\documentclass[useAMS,usenatbib,usegraphicx]{mn2e}
\usepackage{times}
\newcommand{\mycramer}{{Cram\'{e}r}}
\newcommand{\mye}{\mbox{$\mathcal{E}$}}
\newcommand{\myetrue}{\mbox{$\mye_{\rm{true}}$}}
\newcommand{\myb}{\mbox{$\mathcal{B}$}}
\newcommand{\mybtrue}{\mbox{$\mathcal{B_{\rm{true}}}$}}
\newcommand{\mysx}{\mbox{$\mathcal{X}$}}
\newcommand{\mysy}{\mbox{$\mathcal{Y}$}}
\newcommand{\myssigma}{\mbox{$\mathcal{S}$}}
\newcommand{\myuh}{\mbox{$\mathsf{UH}$}}
\newcommand{\mylh}{\mbox{$\mathsf{LH}$}}
\newcommand{\mystep}{\mbox{$\Delta$}}
\newcommand{\mytf}{\mbox{$\mathsf{TF}$}}
\newcommand{\mybf}{\mbox{$\mathsf{BF}$}}
\newcommand{\mytw}{\mbox{$\mathsf{TW}$}}
\newcommand{\mybw}{\mbox{$\mathsf{BW}$}}
\newcommand{\myfwhm}{\mbox{$\mathsf{FWHM}$}}
\newcommand{\mymag}{\mbox{$\mathsf{mag}$}}
\newcommand{\mymagtrue}{\mbox{$\mymag_{\rm{true}}$}}
\newcommand{\myr}{\mbox{$\mathsf{r}$}}

\newcommand{\myvolume}{\mbox{V}}

\newcommand{\myvolumepsf}{\mbox{$V_{\rm{PSF}}$}}
\newcommand{\mysharpness}{\mbox{$\mathsf{sharpness}$}}
\newcommand{\myprf}{\mbox{$\Psi$}} 
\newcommand{\mygprf}{\mbox{$\mathsf{G}$}} 
\newcommand{\mygpsf}{\mbox{$\mathsf{g}$}} 
\newcommand{\mydrf}{\mbox{$\Lambda$}} 
\newcommand{\mybeta}{\mbox{\boldmath$\beta$}}
\newcommand{\mycssl}{\mbox{$\mathcal{L}$}}
\newcommand{\mympd}{\mbox{\large\tt{mpd}}}
\newcommand{\mympdx}{\mbox{\large\tt{mpdx}}}
\newcommand{\myelectrons}{\mbox{e$^-$}}
\newcommand{\myphotons}{\mbox{$\gamma$}}
\newcommand{\mypx}{\mbox{px}}
\newcommand{\myspx}{\mbox{spx}}
\newcommand{\myxi}{\mbox{$x_i$}}
\newcommand{\myyi}{\mbox{$y_i$}}
\newcommand{\myzi}{\mbox{$z_i$}}
\newcommand{\mysnr}{\mbox{$\mathsf{S/N}$}}
\newcommand{\myvarron}{\mbox{$\sigma^2_{\rm{RON}}$}}
\newcommand{\mysigmaron}{\mbox{$\sigma_{\rm{RON}}$}}
\newcommand{\myvarrms}{\mbox{$\sigma^2_{\rm{rms}}$}}
\newcommand{\mysigmarms}{\mbox{$\sigma_{\rm{rms}}$}}

\newcommand{\mymodel}{\mbox{$m$}}
\newcommand{\mymodeli}{\mbox{$m_i$}}
\newcommand{\mybfp}{\mbox{$\mathbf{p}$}}
\newcommand{\mybfr}{\mbox{$\mathbf{r}$}}
\newcommand{\mychi}{\mbox{$\chi^2(\mybfp)$}}
\newcommand{\mybfdelta}{\mbox{\boldmath$\delta$}}
\newcommand{\mybfh}{\mbox{\boldmath$H$}}
\newcommand{\mybfc}{\mbox{\boldmath$C$}}
\newcommand{\myith}{\mbox{$i^{\rm{th}}$}}
\newcommand{\myjth}{\mbox{$j^{\rm{th}}$}}
\newcommand{\myvarebright}{\mbox{$\sigma_{\mbox{\mye:\,bright}}^2$}}
\newcommand{\myvarefaint}{\mbox{$\sigma_{\mbox{\mye:\,faint}}^2$}}
\newcommand{\mysigmae}{\mbox{$\sigma_{\mye}$}}
\newcommand{\mysigmaefaint}{\mbox{$\sigma_{\mbox{\mye:\,faint}}$}}
\newcommand{\mysigmaebright}{\mbox{$\sigma_{\mbox{\mye:\,bright}}$}}
\newcommand{\mysigmab}{\mbox{$\sigma_{\myb}$}}
\newcommand{\mysigmasx}{\mbox{$\sigma_{\mysx}$}}
\newcommand{\mysigmasxbright}{\mbox{$\sigma_{\mbox{\mysx:\,bright}}$}}
\newcommand{\mysigmasy}{\mbox{$\sigma_{\mysy}$}}
\newcommand{\myvarsxbright}{\mbox{$\sigma_{\mbox{\mysx:\,bright}}^2$}}

\newcommand{\myvarsxfaint}{\mbox{$\sigma_{\mbox{\mysx:\,faint}}^2$}}
\newcommand{\myiintinfty}{\mbox{$\displaystyle
\mathop{\int\!\!\!\int}_{\!-\infty}^{\ +\infty}
$}}

\pubyear{accepted May 19, 2005}

\title[Stellar Photometry and Astrometry with Discrete PSFs]{Stellar Photometry
and Astrometry with Discrete Point Spread Functions}

\author[K. J. Mighell]{Kenneth
J. Mighell$^{1}$\thanks{E-mail:  mighell@noao.edu}
  \\
  $^{1}$~National Optical Astronomy Observatory,
  950 North Cherry Avenue, Tucson, AZ~~85719, U.S.A.}

\begin{document}

\date{
}

\pagerange{\pageref{firstpage}--\pageref{lastpage}}

\maketitle

\label{firstpage}

\begin{abstract}
The key features of the MATPHOT algorithm for
precise and accurate stellar photometry and astrometry using
discrete Point Spread Functions are described.
A discrete Point Spread Function (PSF) is a sampled version of a continuous
PSF which describes the two-dimensional probability
distribution of photons from a point source (star) just above the detector.
The shape information about the photon
scattering pattern of a discrete PSF is typically encoded
using a numerical table (matrix) or a FITS image file.
Discrete PSFs are shifted within an
observational model using a 21-pixel-wide damped sinc
function and position partial derivatives are computed using
a five-point numerical differentiation formula.
Precise and accurate stellar photometry
and astrometry is achieved with undersampled CCD observations by using
supersampled discrete PSFs that are sampled 2, 3, or more
times more finely than the observational data.
The precision and accuracy of the MATPHOT algorithm
is demonstrated by using the C-language $\mympd$ code
to analyze simulated CCD stellar observations; measured performance is
compared with a theoretical performance model.
Detailed analysis of simulated
{\sl{Next Generation Space Telescope~}}
observations demonstrate that millipixel relative astrometry
and millimag photometric precision is achievable with complicated
space-based discrete PSFs.
\end{abstract}

\begin{keywords}
techniques: image processing, photometric ---
astrometry ---
instrumentation: detectors ---
methods: analytical, data analysis, numerical, statistical
\vspace*{3truemm}
\end{keywords}

\section{Introduction}

A Point Spread Function (PSF) is a
continuous two-dimensional probability-distribution function
which describes the scattering pattern of photons from a point source (star).

Encoding a PSF as a continuous mathematical function
works well for many ground-based astronomical observations due to
the significant blurring caused by turbulence in the Earth's
atmosphere and dome/telescope seeing.  Ground-based PSFs are
typically characterized by having a lot of the power
in their spatial-frequency distributions at low spatial frequencies.

Space-based PSFs frequently have significant amounts of power at
higher spatial frequencies due to the lack of blurring caused by
atmospheric turbulence.
Adaptive optics can produce PSFs with characteristics
found in both uncorrected ground-based PSFs and space-based PSFs:
low-spatial-frequency features (e.g., broad halos) are frequently
combined with high-spatial-frequency features
(e.g., due to segmented mirrors).

Some PSF-fitting stellar photometric reduction programs describe the PSF
as a combination of continuous mathematical functions and a residual matrix
which contains the difference between the mathematical model of the PSF
and an observed (``true'') PSF.
This artificial breaking of the PSF into
analytical and discrete components is not without mathematical risk.
Such residuals can have small features which are described with
higher spatial frequencies than are present
in the actual observational data --- a problem that can usually be
mitigated by sampling residuals
at higher spatial resolutions than the observational data.

What if we dispose of the use of continuous mathematical
functions to model {\em{any}} part
of the PSF and just use a matrix to describe {\em{all}} of the PSF?
Is precise and accurate stellar photometry and astrometry possible
using matrix PSFs with oversampled stellar image data?
If that is possible, then what extra information, if any,
is required in order to do precision photometric
reductions with matrix PSFs on undersampled data?

This article describes how precise and accurate stellar photometry may
be obtained using PSFs encoded as a matrix.
The following section derives the theoretical performance limits of
PSF-fitting stellar photometry and astrometry. Some of the key features of the
MATPHOT algorithm are presented in \S 3. A demonstration computer program,
called $\mympd$, based on the current implementation of the MATPHOT algorithm,
is described in \S 4.  Simulated CCD (charge-coupled device)
stellar observations are analyzed with
$\mympd$ in \S 5 and the performance of the MATPHOT algorithm is compared with
theoretical expectations.
Concluding remarks are given in \S 6.
An appendix explains box-and-whisker plots which are used extensively
in this article.

\section{Theoretical Performance Limits}

\subsection{Point Response Functions}

A Point Response Function,
$\myprf$,
is the convolution of a
Point Spread Function,
$\phi$,
and a
Detector Response Function,
$\mydrf$
:
\begin{equation}
\myprf
\equiv
\phi\,{\ast}\,\mydrf\ .
\end{equation}
The PSF 
describes the
two-dimensional distribution of {\em{photons}} from a star
{\em{just above the detector.}}
Although stellar photons are distributed as a point source above the
Earth's atmosphere, a stellar image becomes
a two-dimensional distribution
as the stellar photons are scattered by atmospheric turbulence.
The blurred stellar image is then further degraded by
passage of the stellar photons
through the combined telescope and camera optical elements
(such as mirrors, lenses, apertures, etc.).
The PSF is the convolution
of all these blurring effects on the original point-source stellar image.
The two-dimensional discrete (sampled)
Detector Response Function (DRF)
describes how the detector electronics convert stellar
photons ($\myphotons$) to electrons ($\myelectrons$)
--- including such effects as the diffusion of electrons within the detector
substrate or the reflection (absorption) of photons on (in) the gate
structures of the detector electronics.

The PSF is a two-dimensional probability-distribution
function describing the scattering
pattern of a photon. The volume integral of the PSF is one:
$\myvolumepsf \equiv 1$;
photons, after all, have to be scattered {\em{somewhere}}.
It is important to note that since the angular extent of a PSF can be quite
large, the volume integral the PSF {\em{over any given observation}}
is frequently less than one due
to the limited spatial coverage of the observation.

The volume integral of a PRF is, by definition, one or less:
\begin{equation}
\myvolume
\equiv
\myiintinfty
\myprf
\,dx\,dy
=
\myiintinfty
\left(
\phi\,{\ast}\,\mydrf
\right)
\,dx\,dy
{}~~\leq~~1
{}~,
\label{eq:volume}
\end{equation}
where a value that is less than one indicates a loss of stellar photons
during the detection/conversion process within the detector.
While the quantum efficiency (QE) variations within a single detector
are generally not a major problem with state-of-the-art
charge-coupled devices, intrapixel QE variations can be significant
with some near-infrared detector technologies currently being used in
astronomical cameras (e.g., \citealp{lauer99}, \citealp{hookfruchter00}).

A perfect DRF gives a PRF that is a
{\em{sampled version}} of the PSF:
\begin{equation}
\myprf_i
\equiv
\int_{\myxi-0.5}^{\myxi+0.5}
\int_{\myyi-0.5}^{\myyi+0.5}
\phi(x,y)
\,dx\,dy\ ,
\end{equation}
where $i^{\rm{th}}$ pixel of the PRF located at
($\myxi,\myyi$) is the volume integral of the PSF
over the area of the $i^{\rm{th}}$ pixel.
The actual limits of the above volume
integral reflect the appropriate mapping transformation
of the $x$ and $y$ coordinates onto the CCD pixel coordinate system.

The $\mysharpness$ of a PRF is
defined as the volume integral of the {\em{square}}
of the {\em{normalized}}
PRF:
\begin{equation}
\mysharpness
\equiv
\myiintinfty
\tilde{\myprf}^2
\,dx\,dy\
\equiv
\myiintinfty
\left(
\frac{\myprf}{\myvolume}
\right)^2
\,dx\,dy\
\end{equation}
Physically, $\mysharpness$ is a shape parameter which
describes the ``pointiness'' of a PRF;
$\mysharpness$ values range
from a maximum of one
(all of the stellar flux is found within a single pixel)
to a minimum of zero
(a flat stellar image).
For example, cameras that are out of focus have broad PSFs
with $\mysharpness$ values near zero.
A normalized Gaussian PSF
with a standard deviation of $\myssigma$ pixels,
\begin{equation}
\mygpsf\,( x, y; \mysx,\mysy, \myssigma)
\equiv
\frac{1}{2 \pi \myssigma^2}
\exp\left[
-\,\frac{(x-\!\mysx)^{2} \!+ (y-\!\mysy)^{2}}{2 \myssigma^2\!}
\right],
\label{eq:gaussian}
\end{equation}
that has been {\em{oversampled}} with a perfect DRF
will have a $\mysharpness$ value of
\begin{equation}
\myiintinfty
\mygpsf^2( x, y; \mysx,\mysy, \myssigma)
\,\,dx\,dy
=
\frac{1}{4\pi\myssigma^2}\ .
\label{eq:gaussian_sharpness}
\end{equation}
A critically-sampled normalized Gaussian PRF
has a $\mysharpness$ of $1/(4\pi)$
and any PRF with a $\mysharpness$ value
greater than that value ($\sim$$0.0796$)
can be described as being undersampled.
Diffraction limited optics, theoretically,
give $\mysharpness$ values that decrease
(i.e., PSFs become flatter)
with increasing photon wavelength -- for a fixed pixel (detector) size.
With real astronomical cameras, the value of
$\mysharpness$ frequently
depends on
{\em{where the center of a star is located within the central pixel}}
of the stellar image.
For example, the {\sl{Hubble Space Telescope}} {\sl{(HST)}}
WFPC2 Planetary Camera
PRF at a wavelength
of 200 nm has an
observed $\mysharpness$ value
of 0.084 if the PRF is centered in the middle of a PC pixel or
0.063 if the PRF is centered on a pixel corner
\citep[Table 6.5 of][]{biretta_etal2001};
at 600 nm the observed $\mysharpness$ values range from
0.066 (pixel-centered) to 0.054 (corner-centered).
The Wide-Field Cameras of the {\sl{HST}} WFPC2 instrument
have pixels which are approximately half the angular resolution of the
PC camera pixels;
stellar images on the WF cameras are undersampled and
the observed range of WF camera $\mysharpness$ values are
0.102--0.120 at 200 nm and
0.098--0.128 at 600 nm.

The {\em{effective-background area}},
$\mybeta$,
of a PRF
is defined as the {\em{reciprocal}} of the volume integral of the {\em{square}}
of the PRF:
\begin{equation}
\mybeta
\equiv
\left[
\myiintinfty
\myprf^2
\,dx\,dy
\right]^{-1}
\,.
\label{eq:beta}
\end{equation}
Alternatively,
the effective-background area
(a.k.a.
{\em{equivalent-noise area}}
or
{\em{effective solid angle}}
)
of a PRF
is equal to
the reciprocal of the product of
its $\mysharpness$ and the square of its volume:
\begin{equation}
\mybeta
\equiv
\left[
\myiintinfty
\left(
\myvolume
\tilde{\myprf}
\right)^2
\,dx\,dy
\right]^{-1}
\!\!\!=
\,\frac{1}{\myvolume^{\,2}\,\mysharpness}
\,.
\label{eq:beta2}
\end{equation}
The effective-background area of a normalized Gaussian PRF
is $4{\pi}\myssigma^2$ px$^2$,
where $\myssigma$ is the standard deviation in pixels;
a critically-sampled normalized Gaussian PRF
has an effective-background area of $4\pi\approx12.57$ px.
\citet{king83}
notes that numerical integration of a realistic ground-based
stellar profile gives an effective-background area of $30.8\,\myssigma^2$
instead of the value of $4\pi\,\myssigma^2$ for a
normalized Gaussian profile.

\subsection{Basic Least-Squares Fitting Theory}

Consider a CCD
observation of two overlapping stellar images.
Assuming that we already know the PSF and the DRF
of the observation,
a simple model of the observation will have seven parameters:
two stellar intensities\footnote{
Stellar intensity is defined to be
the total number of electrons from a single star
scaled to a PRF volume integral of one.
The {\em{observed}} stellar intensity $(\equiv\mye\myvolume)$
is, by definition, is always less than or equal to
the {\em{measured}} stellar intensity
$(\equiv\mye)\,$.}
$(\mye_1,\mye_2)$ in electrons,
four coordinate values, giving the stellar positions
$(\mysx_1,\mysy_1,\mysx_2,\mysy_2)$ in pixels,
and $\myb$ which is the {\em{observed}} background sky level\footnote{The
observed background sky level (in electrons)
is the product of true background sky level (in photons)
and the {\em{average}} PRF volume {\em{across a pixel}}:
$\myb
\equiv
\mybtrue
\langle\myvolume\rangle$.}
in electrons
(which is assumed to be the same for both stars).
These observational parameters are not independent for overlapping
stars in the presence of photon and CCD readout noise.
The conservation of electron flux will require that if
$\mathcal{E}_1$ increases
then $\mathcal{E}_2$ must decrease and vice versa for a given value of $\myb$.
The most accurate photometry possible is obtained when
these dependent parameters
are fitted simultaneously.  Any reasonable model of two overlapping
stellar images will be a nonlinear function when the positions and
intensities are to be determined simultaneously.  The technique
of nonlinear least-squares fitting was developed to provide for
the simultaneous determination of dependent or independent
parameters of nonlinear model functions.

Assume that we
have a calibrated CCD observation with $N$ pixels
and that $z_i$
is the number of electrons
in the $i^{\rm{th}}$ pixel which is located at the position
of $(x_i,y_i)$ and has a measurement error
of $\sigma_i$ electrons.  Let
$\mymodel( x,y; p_1,\ldots, p_M)$
be an observational model of the CCD electron pixel
values that has two coordinates $(x,y)$ and $M$ parameters.
For notational convenience, let the vector
$\mybfr_i$ represent the coordinates $(x_i,y_i)$ of the $\myith$ pixel
and the vector $\mybfp$ represent all the model parameters
[$\mybfp \equiv (p_1,\ldots,p_M)\,$].
The observational model of the $i^{\rm{th}}$ pixel
can thus be compactly written as follows:
$\mymodeli \equiv \mymodel(\mybfr_i;\mybfp)$.

The measure of the goodness of fit between the
data and the model, called chi square,
is defined as
\begin{equation}
\mychi \equiv \sum_{i=1}^{N}
\frac{1}{\sigma_i^2}
\bigl(\,\myzi~-~\mymodeli\,\bigr)^2\ \ .
\label{eq:chisq}
\end{equation}
The theory of least-squares minimization states that the optimum
value of the parameter vector $\mybfp$
is obtained when $\mychi$
is minimized with respect to each parameter simultaneously.
If $\mybfp_0$ is the optimal parameter vector,
then $\chi^2(\mybfp_0)$ is the absolute minimum
of the $M$-dimensional manifold $\mychi$.

For some small correction parameter vector $\mybfdelta$
one can approximate $\chi^2(\mybfp+\mybfdelta)$ by its Taylor series
expansion:
\begin{eqnarray}
\chi^2\!(\mybfp+\mybfdelta)
&
=
&
\sum_{n=0}^{\infty}
\frac{1}{n!}(\mybfdelta \cdot \nabla)^n \mychi
\nonumber
\\
&
\approx
&
\mychi
{}~+~\mybfdelta\cdot\nabla\mychi
{}~+~\frac{1}{2}\,\mybfdelta\cdot\mybfh\cdot\mybfdelta
\end{eqnarray}
where
\begin{eqnarray}
[\mybfh]_{jk}
&
\equiv
&
\frac{\partial^2\,\mychi}{\partial a_j \partial a_k}
\nonumber
\\
&
\approx
&
\left[\frac{\partial\,\mychi}{\partial a_j}\right]
\left[\frac{\partial\,\mychi}{\partial a_k}\right]
\end{eqnarray}
is the $jk^{\rm{th}}\,$ element of the
$M$$\times$$M$ Hessian matrix $\mybfh$ of $\mychi$
[e.g., \cite{arfken1970}; \cite{press_etal1986}].
The approximation for the calculation of the Hessian matrix elements
is frequently used whenever the computation of the second
partial derivative is numerically unstable.
If $\chi^2(\mybfp+\mybfdelta)$ is a {\em{local}} minimum of $\chi^2$ manifold,
then it can be shown that
\begin{equation}
\mybfh \cdot \mybfdelta = -\nabla\mychi\ .
\end{equation}
By solving this equation for the correction vector $\mybfdelta$\,,
one can determine a better parameter vector as follows:
$
\mybfp^\prime = \mybfp + \mybfdelta\ .
$
When the parameter vector ($\mybfp$) is redefined to be the better
parameter ($\mybfp^\prime$), the Hessian matrix and the
gradient of $\mychi$ can then be recalculated to
determine a new correction vector ($\mybfdelta$).
This process repeats until the correction vector is sufficiently small
-- generally when the difference between solutions is no longer statistically
significant.
If the fitting process has not failed, then the
optimal parameter vector  ($\mybfp_0$)
should be very close to the true parameter vector.

Once the optimal parameter vector has been determined,
the covariance matrix $\mybfc$
may then be calculated by inverting the Hessian matrix $\mybfh$
computed with the optimal parameter vector.
The standard errors (one standard deviation)
of the fitted parameters can be estimated as follows:
\begin{equation}
\sigma_j
\approx
\sqrt{ \left[ \mybfc \right]_{jj} }
=
\left[
\sum_{i=1}^{N}
\frac{1}{\sigma_i^2}
\left(
\frac{\partial \mymodeli}{\partial p_j}
\right)^2\,
\right]^{-1/2}
\label{eq:errorest}
\end{equation}
where $\sigma_j$ is the standard error associated with the
$\myjth$ parameter ($p_j$).  Usage of
equation (\ref{eq:errorest}) for error estimates is based on
the critical assumption that fitted model
parameters are independent
(indicated by negligibly small
off-diagonal elements of the covariance matrix).
It is important to note that whenever this critical assumption is violated,
the results produced by least-squares fitting may not
be statistically reliable, which is to say, they
may no longer be physically meaningful.

\subsection{Photometry}

The theoretical photometric performance limits for
PSF-fitting CCD stellar photometry can be derived using a simple observational
model consisting of a PRF and a constant sky level.

\subsubsection{Observational Model}

Consider a CCD
observation of single isolated star on a flat sky background.
Assuming one already knows the PRF
of the observation at the location of the star,
a simple model of the observation would have just two parameters:
the stellar intensity ($\mye$) in electrons,
and the observed background sky level ($\myb$) in electrons.
The observational model for the $i^{\rm{th}}$ pixel would be
\begin{equation}
\mymodeli
\equiv
\myb
+
\mye
\myvolume \tilde{\myprf}_i
\ ,
\label{eq:photmodel}
\end{equation}
where $\myvolume$ is the volume integral
of the PRF
and $\tilde{\myprf}_i$ is the value of the $\myith$
pixel of the {\em{normalized}} PRF
$(\,\tilde{\myprf}_i \equiv \myprf_i / \myvolume\,)$.

\subsubsection{Bright Star Limit}

In the case of bright stars, most of the electrons found in the $\myith$
pixel of the observation will come from the star and not the sky:
\begin{equation}
\mymodeli \approx \mye \myvolume \tilde{\myprf}_i
\ .
\end{equation}
The actual number of electrons found in the $\myith$ pixel will be
described by a Poisson distribution with a mean and variance of
$\mymodeli$.  The measurement error (one standard deviation)
for the $\myith$ pixel would
thus be
\begin{eqnarray}
\sigma_i
&
=
&
\sqrt{ \,\mymodeli }
\nonumber
\\
&
\approx
&
\sqrt{ \mye \myvolume \tilde{\myprf}_i }
\label{eq:bphoterr}
\ .
\end{eqnarray}
All other noise sources
(due to, for example, the observed background sky,
instrumental readout noise, flat-field
calibrations errors, etc.)
are assumed, in this special case, to be negligibly small.

The variance
of the stellar intensity measurement error of bright
stars can be estimated
using equations
(\ref{eq:errorest}),
(\ref{eq:photmodel}),
and
(\ref{eq:bphoterr}):
\begin{eqnarray}
\myvarebright
&
\approx
&
\left[
\sum_{i=1}^{N}
\frac{1}{\mye\myvolume\tilde{\myprf}_i}
\left(
\frac{\partial}{\partial \mye}
\mye \myvolume \tilde{\myprf}_i
\right)^2\,
\right]^{-1}
\nonumber
\\
&
\approx
&
\frac{\mye}{\myvolume}
\left[
\myiintinfty
\tilde{\myprf}
\,dx\,dy
\right]^{-1}
\nonumber
\\
&
=
&
\frac{\mye}{\myvolume}
\ ,
\label{eq:myvarebright}
\end{eqnarray}
as expected from photon statistics.

A bright isolated star with an intensity of $10^6$ photons
imaged with a perfect CCD detector
would have a stellar image
with $10^6$ $\myelectrons$ $(=\mye)$
and a stellar intensity measurement error
of $\mysigmae \approx \sqrt{\mye/(\myvolume\!\equiv\!1)} = 10^3$
$\myelectrons$.
The same star imaged with an inefficient CCD detector with a quantum
efficiency of 25\% ($\myvolume=1/4$) would have a stellar image
with $\sim$$250,000$ $\myelectrons$ which
would have a Poisson noise error of $\sim$$500$ $\myelectrons$.
The {\em{measured}} stellar intensity
is $\mye\approx10^6$ $\myelectrons$ with an rms measurement error of
$\mysigmae \approx \sqrt{\mye/\myvolume} = 2000$ $\myelectrons$
which is two times larger than it would be with
a perfect detector and
four times larger than the Poisson noise error
of the {\em{observed}} number of electrons.

Solving for {\em{measured}} stellar intensity $(\equiv\mye)$
instead of the {\em{observed}} stellar intensity $(\equiv\mye\myvolume)$
enables the creation of stellar photometric reduction programs capable of
dealing with intrapixel QE variations through the accurate
modeling of the image formation process within the detector.
While it is certainly convenient to assume that one's detector has negligible
intrapixel QE variation, in the real world even NASA-grade CCD detectors,
like those found in the {\sl{HST}} WFPC2 instrument,
can have peak-to-peak intrapixel
sensitivity variations that are greater than 0.02 mag ($>$2\%)
\citep[see Figs.\ 5 and 6 of][]{lauer99}.

\subsubsection{Faint Star Limit}

Most of the electrons found in the $\myith$ pixel of an observation
of a {\em{faint}} isolated star on a flat sky background
will come from the sky and not from the star.
In that case, the measurement error
associated with
$\myith$ pixel is approximately the
effective-background noise level:
\begin{equation}
\sigma_i
\approx
\mysigmarms\,,
\label{eq:fphoterrirms}
\end{equation}
where
\begin{eqnarray}
\mysigmarms
&
\equiv
&
\sqrt{\displaystyle\frac{1}{N}\sum_{i=1}^{N} \sigma_i^{2}}
\label{eq:fphoterr}
\\
&
\approx
&
\sqrt{ {\displaystyle\myb} + \myvarron }\ ,
\label{eq:fphoterr_approx}
\end{eqnarray}
$\myb$ is the constant observed background sky level
which is assumed to be a Poisson distribution with a mean of $\myb$ electrons,
and
$\mysigmaron$\ is
the rms readout noise.

The variance of the stellar intensity measurement
error of faint stars can be estimated
using equations
(\ref{eq:errorest}),
(\ref{eq:photmodel}),
(\ref{eq:fphoterrirms}),
(\ref{eq:fphoterr}),
(\ref{eq:fphoterr_approx}),
and
(\ref{eq:beta2}):
\begin{eqnarray}
\myvarefaint
&
\approx
&
\left[
\sum_{i=1}^{N}
\frac{1}{\myvarrms}
\left(
\frac{\partial}{\partial \mye}
\mye \myvolume \tilde{\myprf}_i
\right)^2\,
\right]^{-1}
\nonumber
\\
&
\approx
&
\frac{\myvarrms}{\myvolume^{\,2}}
\left[
\myiintinfty
\tilde{\myprf}^2
\,dx\,dy
\right]^{-1}
\nonumber
\\
&
=
&
\mybeta\,\myvarrms
\\
&
\approx
&
\mybeta\,
\left[
\myb + \myvarron
\right]
\ ,
\label{eq:myvarefaint}
\end{eqnarray}
where $\mybeta$ is the
effective-background area of the PRF.
Equation (\ref{eq:myvarefaint}) agrees
with equation (9) of \citet{king83} for
a perfect $(\myvolume\!\equiv\!1)$
noiseless $(\mysigmaron\!\equiv\!0~\myelectrons)$
detector.

An important additional noise source for the photometry of faint stars
is the systematic error due to the
uncertainty of the measurement of the background.
If the sky background is assumed to be flat, then
the rms measurement error
of the constant sky background
can be estimated
using equations
(\ref{eq:errorest}),
(\ref{eq:photmodel}),
(\ref{eq:fphoterrirms}),
(\ref{eq:fphoterr}),
and
(\ref{eq:fphoterr_approx}):
\begin{eqnarray}
\mysigmab
&
\approx
&
\left[
\sum_{i=1}^{N}
\frac{1}{\myvarrms}
\left(
\frac{\partial}{\partial \myb}
\myb
\right)^2\,
\right]^{-1/2}
\nonumber
\\
&
=
&
\frac{\mysigmarms}{\sqrt{N}}
\\
&
\approx
&
\sqrt{
\frac{\myb + \myvarron}{N}
}
\ .
\label{eq:sigmab}
\end{eqnarray}
Given a CCD observation with no readout noise, equation (\ref{eq:sigmab})
reduces to the value of $\mysigmab = \sqrt{\myb/N}$
expected from simple sampling statistics.

The portion of the rms stellar intensity measurement error
that is caused by the error in the determination
of the local sky level is
$\mysigmab\,\mybeta$ \citep{irwin85}.
While this error is frequently negligible for bright stars,
it is generally significant for faint stars.
Including the uncertainty in the determination of the constant
observed background sky level thus
gives a more realistic estimate for the
rms stellar intensity measurement error
for {\em{faint}} stars:
\begin{eqnarray}
\mysigmaefaint
&
\approx
&
\sqrt{\mybeta\myvarrms}
+
\mybeta \mysigmab
\nonumber
\\
&
=
&
\sqrt{\mybeta}
\left(
1
+
\sqrt{\mybeta/N}
\,\right)
\mysigmarms
\\
&
\approx
&
\sqrt{\mybeta}
\left(
1
+
\sqrt{\mybeta/N}
\,\right)
\sqrt{
\myb + \myvarron
}
\ .
\label{eq:myrmsefaint}
\end{eqnarray}
Precise and accurate stellar photometry of faint stars requires
an excellent determination of the observed background sky which in turn
requires accurate background sky models.
Given a valid background sky model,
small apertures will be more sensitive to background sky measurement errors
than large apertures.

\subsubsection{Photometric Performance Model}

A realistic
photometric performance model for CCD PSF-fitting photometry can be
created by combining
the bright and faint star limits developed above.
The theoretical {\em{upper limit}} for
the photometric signal-to-noise ratio ($\mysnr$)
of CCD PSF-fitting photometric algorithms is
as follows:
\begin{eqnarray}
\mysnr
&
\equiv
&
\frac{\mye}{\mysigmae}
\nonumber
\\
&
\approx
&
\frac{\mye}{
\sqrt{
\myvarebright
+
\myvarefaint
}
}
\nonumber
\\
&
\approx
&
\frac{\mye}{
\sqrt{
\displaystyle\frac{\mye}{\myvolume}
+
\mybeta
\left(1 + \sqrt{\mybeta/N}\,\right)^2
\myvarrms
}
}
\label{eq:theoretical-snr}
\\
&
\approx
&
\frac{\mye}{
\sqrt{
\displaystyle\frac{\mye}{\myvolume}
+
\mybeta
\left(1 + \sqrt{\mybeta/N}\,\right)^2
\left[
\myb + \myvarron
\right]
}
}
\ .
\label{eq:practical-snr}
\end{eqnarray}
These approximations assume,
for the sake of simplicity,
that any noise contribution
due to dark current
and quantization noise
is negligible.
While these additional noise sources can be added
to create an even more
realistic performance model for stellar photometry,
the assumption of low dark current and minimal quantization noise
is realistic for state-of-the-art
astronomical-grade CCD imagers.
The resulting photometric error is approximately
\begin{equation}
\Delta \mymag
\approx
\frac{1.0857}{\mysnr}
\ ,
\end{equation}
where the constant $1.0857$ is an approximation for Pogson's ratio
$a\!\equiv\!5/\ln(100)\!=\!2.5\log(e)\,$
\citep{pogson1856}.

\subsubsection{\mycramer-Rao Lower Bound}

The \mycramer-Rao Lower Bound (CRLB) is the lower bound on the variance
of {\em{any}} unbiased estimator.
Since
it is physically impossible to find an unbiased estimator that beats the CRLB,
the CRLB provides a performance benchmark against which any unbiased estimator
can be compared.

The \mycramer-Rao Lower Bound for stellar photometry
of a single isolated star imaged by a two-dimensional
photon-counting detector
has been derived several times in the astrophysical literature
(see, e.g., Appendix A of \citealt{perryman_etal:1989},
\citealt{irwin85},
and \citealt{king83}).
The generalization for a crowded field with overlapping stellar images
is given in \cite{jakobsen_etal:1992}.

The \mycramer-Rao Lower Bound for the bright star limit of
stellar photometry of a single isolated star is
\begin{equation}
\sigma_{\mbox{\mye:\,bright-CRLB}}^2 = \mye
\end{equation}
which is equation (\ref{eq:myvarebright})
with a perfect detector.
The CRLB for the faint star limit of
stellar photometry of a single isolated star is
\begin{equation}
\sigma_{\mbox{\mye:\,faint-CRLB}}^2 = \mybeta\,\myb
\end{equation}
which is equation (\ref{eq:myrmsefaint})
with a noiseless detector
and a negligible background measurement error
($N\!\rightarrow\!\infty$).

The photometric performance model has bright and faint star limits
which are the same, respectively, as the
bright and faint star \mycramer-Rao Lower Bounds for stellar photometry
of a single isolated star on a flat sky background
imaged with a perfect noiseless detector.

\subsection{Astrometry}

The theoretical astrometric limits for
PSF-fitting CCD stellar photometry can be derived using a simple observational
model consisting of a Gaussian PRF and a constant sky level.

\subsubsection{Observational Model}

Consider a CCD
observation of single isolated star on a flat sky background.
A Gaussian is a good model for
the PSF of a ground-based CCD observation
since the central core of a ground-based stellar profile is approximately
Gaussian \citep{king71}.
In this case the PSF would have three parameters:
two coordinate values giving the location ($\mysx,\mysy$) of the star
on the CCD
and the standard deviation of the Gaussian ($\myssigma$) in pixels
[see equation (\ref{eq:gaussian})].

An {\em{imperfect but uniformly flat}} DRF ($\myvolume\!<\!1$)
gives a value for the $\myith$ pixel of the PRF located at $(\myxi,\myyi)$
which is equal to the product of
the volume of the PRF
and the value of the
volume integral of the PSF over the area of the $\myith$ pixel:
\begin{equation}
\mygprf_i
\equiv
\myvolume
\int_{\myxi-0.5}^{\myxi+0.5}
\int_{\myyi-0.5}^{\myyi+0.5}
\mygpsf\,( x, y; \mysx,\mysy, \myssigma)
\,dx\,dy\ ,
\end{equation}
The actual limits of the above volume
integral reflect the appropriate mapping transformation
of the $x$ and $y$ coordinates onto the CCD pixel coordinate system.

If the PRF has been {\em{oversampled}}, the
value of the $\myith$ pixel of the PRF
is approximately equal to the product of
the volume of the PRF
and
the value of the PSF at the {\em{center of the $\myith$ pixel}}:
\begin{equation}
\mygprf_i
\approx
\myvolume\,\mygpsf_i
\label{eq:gprfi_approximation}
\end{equation}
where
\begin{equation}
\mygpsf_i
\equiv
\mygpsf\,( \myxi, \myyi; \mysx,\mysy, \myssigma)~.
\end{equation}

A simple model of the observation will require two additional parameters:
the stellar intensity ($\mye$) in electrons
and the observed background sky level ($\myb$) in electrons.
The $\myith$ pixel of the observational model
would be
\begin{equation}
\mymodeli
\equiv
\myb
+
\mye
\myvolume
\tilde{\mygprf}_i
\ ,
\label{eq:astmodel}
\end{equation}
where $\myvolume$ is the volume integral
of the PRF
and $\tilde{\mygprf}_i$ is the value of the $\myith$
pixel of the {\em{normalized}} PRF
$(\,\tilde{\mygprf}_i \equiv \mygprf_i / \myvolume \approx \mygpsf_i\,)$.

\subsubsection{Bright Star Limit}

In the case of bright stars, most of the electrons found in the $\myith$
pixel of the observation will come from the star and not the sky:
\begin{equation}
\mymodeli \approx \mye \myvolume \tilde{\mygprf}_i
\ .
\end{equation}
The actual number of electrons found in the $\myith$ pixel will be
described by a Poisson distribution with a mean and variance of
$\mymodeli$.  The measurement error (one standard deviation)
for the $\myith$ pixel would
thus be
\begin{eqnarray}
\sigma_i
&
=
&
\sqrt{ \,\mymodeli }
\nonumber
\\
&
\approx
&
\sqrt{ \mye \myvolume \tilde{\mygprf}_i }
\label{eq:abphoterr}
\ .
\end{eqnarray}
All other noise sources
(e.g., due to the observed background sky,
instrumental readout noise, flat-field
calibrations errors, etc.)
are assumed to be negligibly small.

The variance
of the stellar $\mysx$ position measurement error
of a bright isolated {\em{oversampled}} star
can be estimated
using equations
(\ref{eq:errorest}),
(\ref{eq:astmodel}),
(\ref{eq:abphoterr}),
and
(\ref{eq:gaussian}):
\begin{eqnarray}
\mbox{$\myvarsxbright\!\!\!\!\!\!\!\!\!\!\!\!\!\!\!\!\!\!\!$}
\nonumber
\\
&
\approx
&
\left[
\sum_{i=1}^{N}
\frac{1}{\mye\myvolume\tilde{\mygprf}_i}
\left(
\frac{\partial}{\partial \mysx}
\mye \myvolume \tilde{\mygprf}_i
\right)^2\,
\right]^{-1}
\nonumber
\\
&
\approx
&
\frac{1}{\mye \myvolume}
\left[
\sum_{i=1}^{N}
\frac{1}{\mygpsf_i}
\left(
\frac{\partial}{\partial \mysx}
\mygpsf_i
\right)^2\,
\right]^{-1}
\nonumber
\\
&
\approx
&
\frac{\myssigma^4}{\mye \myvolume}
\left[
\myiintinfty
\mygpsf( x, y; \mysx, \mysy, \myssigma)
\,(x-\mysx)^2
\,dx\,dy
\right]^{-1}
\label{eq:sxbt1}
\nonumber
\\
&
=
&
\frac{\myssigma^2}{\mye\myvolume}
\nonumber
\\
&
\approx
&
\frac{\mycssl^2}{\mye\myvolume}
\,,
\label{eq:sxb_cssl}
\end{eqnarray}
where
\begin{equation}
\mathcal{L}
\equiv
\sqrt{\frac{\mybeta\,\myvolume^2}{4\pi}}
\label{eq:cssl}
{}~=~
\frac{1}{\sqrt{4\pi\,\mysharpness}}
\end{equation}
is the {\em{critical-sampling scale length}} of the PRF\footnote{From the
definition of the effective-background area of an oversampled Gaussian PRF
with $\myvolume\!<\!1$,
$\,\mybeta_{\mathsf{G}}
\equiv
4{\pi}\myssigma^2/\myvolume^2\,$,
one sees that critical-sampling scale length has been designed to be
a proxy for $\myssigma$ for {\em{any}} PRF.
} in pixel units ($\mypx$), which, unlike $\myssigma$, is defined for all PRFs.
By definition, the critical-sampling scale length of a
critically-sampled PRF imaged with
a perfect detector is one pixel;
$\mycssl > 1$ indicates that the PRF is {\em{oversampled}}, while
$\mycssl < 1$ indicates that the PRF is {\em{undersampled}}.

In the special case of a critically-sampled
bright star imaged with a perfect detector,
one finds that the astrometric performance limit (in pixel units)
is equal to the reciprocal of photometric error performance limit:
\begin{displaymath}
\mysigmasxbright
\approx
\frac{1}{\sqrt{\mye}}
\approx
\frac{1}{\mysigmaebright}
\,.
\end{displaymath}

\subsubsection{Faint Star Limit}

Let us again assume that the noise contribution from the star
is negligibly small and that the variance of the measurement
error of the $\myith$ pixel can be replaced with an average
constant rms value.
The variance
of the stellar $\mysx$ position measurement error
of a faint isolated {\em{oversampled}} star
can be estimated using
equations
(\ref{eq:errorest}),
(\ref{eq:astmodel}),
(\ref{eq:fphoterrirms}),
(\ref{eq:fphoterr}),
(\ref{eq:fphoterr_approx}),
and
(\ref{eq:gaussian}):
\begin{eqnarray}
\mbox{$\myvarsxfaint\!\!\!\!\!\!\!\!\!\!\!\!\!\!\!\!\!\!\!$}
\nonumber
\\
&
\approx
&
\left[
\sum_{i=1}^{N}
\frac{1}{\myvarrms}
\left(
\frac{\partial}{\partial \mysx}
\mye \myvolume \tilde{\mygprf}_i
\right)^2\,
\right]^{-1}
\nonumber
\\
&
\approx
&
\frac{\myvarrms}{\mye^2\myvolume^{\,2}}
\left[
\sum_{i=1}^{N}
\left(
\frac{\partial}{\partial \mysx}
\mygpsf_i
\right)^2\,
\right]^{-1}
\nonumber
\\
&
\approx
&
\frac{\myvarrms\,\myssigma^4}{\mye^2\myvolume^{\,2}}
\left[
\myiintinfty
\mygpsf^2( x, y; \mysx,\mysy, \myssigma)
\,\,(x-\mysx)^2
\,dx\,dy
\right]^{-1}
\nonumber
\\
&
=
&
8\pi
\,\myvarrms
\frac{\myssigma^4}{\mye^2\myvolume^{\,2}}
\nonumber
\\
&
\approx
&
8\pi
\,\myvarrms
\left(
\frac{\mathcal{L}^2}{\mye\myvolume}
\right)^{\!2}
\nonumber
\\
&
\approx
&
8\pi
\,\myvarrms
\left(
\myvarsxbright
\right)^2
\\
&
\approx
&
8\pi
\left(
\myb + \myvarron
\right)
\left(
\myvarsxbright
\right)^2
\label{eq:sxf_cssl_approx}
\,.
\end{eqnarray}

\bigskip
\subsubsection{Astrometric Performance Model}

A realistic
performance model for CCD PSF-fitting astrometry can be
created by combining
the bright and faint star limits developed above.
The expected {\em{lower limit}} of the
rms measurement error for the stellar
$\mysx$ position for a
single isolated star on a flat sky can be estimated as follows:
\begin{eqnarray}
\mysigmasx
&
\approx
&
\sqrt{
\myvarsxbright
+
\myvarsxfaint
}
\nonumber
\\
&
\approx
&
\sqrt{
\frac{\mathcal{L}^2}{\mye\myvolume}
\,
\left[
1
+
8\pi\,\myvarrms
\frac{\mathcal{L}^2}{\mye\myvolume}
\,
\right]
}
\\
&
\approx
&
\sqrt{
\frac{\mathcal{L}^2}{\mye\myvolume}
\,
\left[
1
+
8\pi
\left(
\myb + \myvarron
\right)
\frac{\mathcal{L}^2}{\mye\myvolume}
\,
\right]
}
\,.
\label{eq:practical-astrometric-model-x}
\end{eqnarray}
The rms stellar $\mysy$ position measurement error is, by symmetry, the same
as for $\mysx\,$:
\begin{equation}
\mysigmasy
=
\mysigmasx
\label{eq:practical-astrometric-model-y}
\ .
\end{equation}

\subsubsection{Photonic Limit and the \mycramer-Rao Lower Bound}

The \mycramer-Rao Lower Bound for stellar astrometry depends not only on
the signal-to-noise ratio
but also on the {\em{size}} and {\em{shape}} of the detector.
For well-sampled data, the size and shape of the detector can be ignored
and a CRLB
can be found for a perfect noiseless detector with infinitely
small pixels. This is called the {\em{photonic limit}}.

The determination of the CRLB for astrometry becomes
much more complicated with {\em{undersampled}} observations.
Astrometric precision degrades
when the size of the detector is comparable to the size of the stellar
image -- the quality of the position estimation is then dependent
on the fraction of photons falling {\em{outside}} of the central pixel.
The worst-case scenario for stellar astrometry
occurs when {\em{all}} of the light from a star falls
within a single pixel:
all one knows for sure, in that unfortunate case,
is that the star is located {\em{somewhere}} within the
central (and only) pixel.

The photonic limit (PL) for
stellar astrometry of a bright well-sampled
single isolated normalized Gaussian star is
\begin{displaymath}
\sigma_{\mbox{\mysx:\,bright-PL}}^2
=
\frac{\myssigma^2}{\mye}
\end{displaymath}
\citep{irwin85}.
Using $\mycssl$ as a proxy for $\myssigma$, one has
the generalized form for any PSF:
\begin{equation}
\sigma_{\mbox{\mysx:\,bright-PL}}^2
\approx
\frac{\mycssl^2}{\mye}
\,,
\end{equation}
which is equation (\ref{eq:sxb_cssl}) with a perfect detector.

The photonic limit for
stellar astrometry of a faint well-sampled
single isolated normalized Gaussian star is
\begin{displaymath}
\sigma_{\mbox{\mysx:\,faint-PL}}^2
=
\frac{8\pi\,\myb\,\myssigma^4}{\mye^2}
\end{displaymath}
\citep{irwin85}.
Using $\mycssl$ as a proxy for $\myssigma$, one has
the generalized form for any PSF:
\begin{equation}
\sigma_{\mbox{\mysx:\,faint-PL}}^2
\approx
\frac{8\pi\,\myb\,\mathcal{L}^4}{\mye^2}
\,,
\end{equation}
which
is equation (\ref{eq:sxf_cssl_approx}) with a perfect noiseless detector.

The astrometric performance model has bright and faint star limits
which are the same, respectively, as the
bright and faint star photonic astrometric limits, which are
the \mycramer-Rao Lower Bounds for stellar astrometry
of a single isolated Gaussian star on a flat sky background
imaged with a perfect noiseless detector with infinitely small pixels.
The \mycramer-Rao Lower Bound for stellar astrometry
of a single isolated Gaussian star on a flat sky background
imaged with a perfect noiseless CCD {\em{with square pixels}}
\citep{winick:1986} quickly approaches the photonic limits
with {\em{well-sampled}} observations; undersampled observations will have
larger astrometric errors than predicted by the photonic limits.

\subsection{Relation between Astrometric and Photometric Errors}

\subsubsection{Bright Star Limit}

Following \citet{king83} and \citet{irwin85},
I now compare the astrometric error
of bright isolated stars with their photometric error.
The ratio of the astrometric error of a bright isolated star
and the critical-sampling scale length
of the PRF is
equal to the ratio of the stellar intensity measurement error
and the stellar intensity:
\begin{equation}
\frac{\sigma_{\mathcal{X}}}{\mathcal{L}}
=
\frac{\sigma_{\mathcal{E}}}{\mathcal{E}}
\label{eq:awrtpbsl}
\ .
\end{equation}
For example, a bright isolated critically-sampled
star with one million
electrons imaged on a perfect detector
($\mathcal{E}\!=\!10^6$ \myelectrons,
$V\!\equiv\!1\,$,
and $\mycssl\!=\!1$ \mypx\,)
would, theoretically, have a signal-to-noise ratio of a thousand
($\mysnr\!=\!1000$),
a stellar intensity measurement error of
$\mysigmae\!=\!1000$ \myelectrons,
and an rms position error in $x$
of one-thousandth of a pixel $(\mysigmasx\!=\!0.001~\mypx\,)$.
Such astrometric accuracy may be difficult to achieve in practice
under normal ground-based observing conditions even with state-of-the-art
astronomical-grade CCD cameras.
\vspace*{-2truemm}

\subsubsection{Faint Star Limit}

The astrometric error
of faint isolated stars is related to their photometric error as follows:
\begin{equation}
\frac{\sigma_{\mathcal{X}}}{\mathcal{L}}
\approx
\left(
\frac{
\sigma_{\mathcal{E}}
}{
\mathcal{E}
}
\right)
\frac{\sqrt{2}}{1+\sqrt{\beta/N}}
\ .
\end{equation}
For example, a faint isolated critically-sampled star
imaged with a perfect detector
with a 20.0\% intensity measurement error
and a negligible background measurement error
($N\!\rightarrow\!\infty$)
would, theoretically, have an
astrometric error of
$\sim$0.283 $\left[\approx\!(0.200)\,\sqrt{2}\,\right]$ px\,.
\vspace*{-2truemm}

\subsubsection{Practical Lower Bound}

These results suggest the following practical lower bound
for astrometric errors with respect to photometric errors:
\begin{center}
\parbox{0.40\textwidth}{\em{$X\,$\% photometry gives no
better than $X\,$\% astrometry\break with respect to the
critical-sampling scale length ($\mycssl$)\,.
}}
\end{center}
For example, a star with one-percent
stellar photometry will have no better than
one-percent astrometry with respect to the critical-sampling scale length.
If the star is critically sampled,
then the astrometric precision will be no better than 0.01 px.

All of the above derivations are based on the assumption that
that flat-field calibration errors are negligible.
The relation between photometry and astrometry
for bright isolated stars
can fail with large flat-field calibration errors.

\section{Discrete Point Spread Functions}

A {\em{discrete}} Point Spread Function is
a sampled version of a continuous two-dimensional Point Spread Function.
The shape information about the photon scattering pattern of a discrete PSF
is typically encoded using a numerical table (matrix).
An {\em{analytical}} PSF has the shape information encoded with
continuous two-dimensional mathematical functions.

In order to do accurate stellar photometry and astrometry with discrete PSFs
one needs to able to
(1) accurately shift discrete PSFs to new positions
within the observational model,
and (2) compute the position partial derivatives of discrete PSFs.
The next two
subsections describe how these tasks may be accomplished using
numerical analysis techniques.

\subsection{Moving Discrete PSFs}

Building a realistic observation model requires the placement
of a star at the desired location within the model; this is done
by determining the PRF at required location and then multiplying
it by the stellar intensity.
With PSFs encoded by mathematical functions,
one just computes the PSF at the desired location in the observational
model.  With discrete PSFs,
one ideally takes a reference PSF
(typically derived/computed for the center of a pixel)
and shifts it to the desired location using a perfect two-dimensional
interpolation function.
But how is this done in practice?
The sinc function,
$\sin(\pi x)/(\pi x)$,
is, theoretically, a perfect two-dimensional interpolation function.
Unfortunately, the sinc function decays with $1/x$ and never actually
reaches zero.
One can use a windowed interpolant
in order to improve computational speed
--- but one must be cautious
about aliasing effects caused by using a windowed function.
In the case of stellar photometry and astrometry, {\em{aliasing effects will
generally only be seen with bright stars}} since a large number of
photons are required in order to adequately sample the higher spatial
frequencies of the PSF.

The following 21-pixel-wide
{\em{damped sinc function}} interpolant
does an excellent job interpolating discrete PSFs:
\begin{eqnarray}
\mbox{$f^{\rm{shifted}}(x_0)
\!\!\!\!\!\!\!\!\!\!\!\!\!\!\!\!\!\!\!\!\!\!\!\!\!\!\!\!$}
\nonumber
\\
&
\equiv
&
\!\!\!\!\!
\sum_{i=-10}^{10}
\!\!f(x_i)
\frac{\sin\left(\pi(x_i-x_0)\right)}{\pi(x_i-x_0)}
\exp{\left(-\left[\frac{x_i-x_0}{3.25}\right]^2\right)}
\label{eq:dampedsinc}
\end{eqnarray}
\noindent
\citep{mighell_spie:2002}.
Note that since the two-dimensional sinc function is separable in $x$ and $y$,
this interpolant can be coded to be computationally fast and efficient.
This interpolant,
from the ZODIAC C library written by Marc Buie of Lowell Observatory,
was specifically designed for use with 32-bit floating numbers.

Aliasing problems due to critically-sampled or undersampled data
may be overcome by using discrete PSFs that are {\em{super\-sampled}}
at 2, 3, or more times more finely than the observational data.
In order to have a realistic observational model,
once the supersampled discrete PSF has been interpolated to the correct
position, a new {\em{degraded}} (rebinned) version of the discrete PSF
must be created which has the same spatial resolution as
the observational data.

\subsection{Position Partial Derivatives of Discrete PSFs}

While the mathematics of determining the position partial derivatives
of individual stars within
the observational model with respect to the $x$ and $y$ direction vectors
is the same regardless of how the shape information in a PSF is encoded,
the implementation methodology for the computation of
position partial derivatives of discrete PSFs is very different than
the one used for analytical PSFs.

The position partial derivatives of discrete PSFs can be determined using
{\em{numerical differentiation techniques}} on the discrete PSF.

It is a standard practice in numerical analysis to approximate the
first, second, or higher, derivatives of a tabulated function $f(x_i)$
with multi-point formulae.
\cite{abramowitz_stegun:1964}
give 18 different multi-point formulae which can be used
(with varying degrees of accuracy) to
approximate the first derivative of the tabulated
function $f(x_i)$.
The following
five-point differentiation formula,
\begin{eqnarray}
\mbox{$f^{\,\prime}(x_i)\!\!\!\!\!\!\!\!\!\!\!\!\!\!\!$}
\nonumber
\\
&
\approx
&
\!\!\!
\frac{1}{12}\left[
f(x_{i-2})
- 8\,f(x_{i-1})
+ 8\,f(x_{i+1})
- f(x_{i+2})
\right]\,,
\label{eq:five-point}
\end{eqnarray}
\noindent
\citep[p.~914 of][]{abramowitz_stegun:1964}
works well with discrete PSFs
\citep{mighell_spie:2002}.
This approximation
takes just
4 additions
and
3 multiplications
which generally makes it considerably faster
to compute than the traditional determination
of the partial derivative of the volume integral of the PSF above a CCD
pixel.

\section{The MATPHOT Algorithm}

The concepts presented above outline the unique and fundamental
features of the MATPHOT algorithm for accurate and precise stellar
photometry using discrete Point Spread Functions.

While the key features of a CCD stellar photometric reduction algorithm
can be described in an article, the full implementation of such an
algorithm generally exists as a complex computer program consisting
of many thousands of lines of computer code.
Since good algorithms can be poorly implemented,
it can be difficult to differentiate between
a poor algorithm and a poorly-coded implementation of a good algorithm.

Confidence in a complex algorithm can be established by
developing an implementation of the algorithm
which meets theoretical performance expectations.
The following subsection
describes a real-world implementation of the MATPHOT algorithm that
meets the theoretical performance expectations
for accurate and precise stellar photometry and astrometry
which were derived in \S 2.

\subsection{MPD: MatPhot Demonstrator}

I have written a C-language computer program,
called $\mympd$\footnote{All
source code and documentation for $\mympd$ and support software
is freely available at the official MATPHOT website at NOAO:
{}~~~http://www.noao.edu/staff/mighell/matphot},
which is based on the current
implementation of the MATPHOT algorithm
for precise and accurate stellar photometry using
discrete Point Spread Functions.
The $\mympd$ code demonstrates the precision and accuracy of
the MATPHOT algorithm by
analyzing simulated CCD observations based on user-provided discrete PSFs
encoded as FITS images \citep{wells_etal:1981}.
Discrete PSFs are shifted within
the observational model using the 21-pixel-wide damped sinc
interpolation function given in equation (\ref{eq:dampedsinc}).
Position partial derivatives of discrete PSFs are computed
using the five-point differentiation formula given in
equation (\ref{eq:five-point}).
Accurate and precise stellar photometry and astrometry of
{\em{undersampled}} CCD observations can be obtained with the
$\mympd$ code when it is presented with
{\em{supersampled}} discrete PSFs that are sampled 2, 3, or more times
more finely than the observational data.
The $\mympd$ code is based on a robust implementation of
the Levenberg-Marquardt method of nonlinear least-squares minimization
(\citealp{levenberg:1944},
\citealp{marquardt:1963},
also
\citealp{mighell:1989}).
When presented with simulated observations based on a Gaussian PSF
with a known $\myfwhm$ value\footnote{The $\myfwhm$
(Full-Width-at-Half-Maximum) value of a Gaussian is equal to
$2\sqrt{\ln(4)}$ times the standard deviation, $\myssigma$, of the Gaussian:
$\myfwhm \approx 2.35482\,\myssigma$
[see equation (\ref{eq:gaussian})].},
the $\mympd$ code can analyze the observation two different ways:
(1) the MATPHOT algorithm can be used with a discrete Gaussian PSF,
or (2) analytical techniques \citep{mighell:1989,mighell:1999}
can be used with an analytical Gaussian PSF.

\section{Simulated Observations}

\subsection{Oversampled PSFs}

I now demonstrate that the theoretical performance limits of \S 2
provide practical performance metrics for photometry and astrometry
of CCD stellar observations that are analyzed with oversampled
Gaussian Point Spread Functions.

\subsubsection{Analytical PSFs}

Twenty thousand {\em{oversampled}} CCD
stellar observations were simulated and analyzed using the
$\mympd$ code.
The CCD detector was assumed to be perfect ($\myvolume\!\equiv\!1$)
with a CCD readout noise value of
$\mysigmaron=3$ $\myelectrons\,\mypx^{\,-1}\,$.
Stars were simulated using an {\em{analytical}} Gaussian PSF with a
$\myfwhm\equiv3$ $\mypx$ located near the center of 60$\times$60 pixels,
the input stellar intensities ranged from
$-6$ to $-15$ mag\footnote{The MATPHOT magnitude system assumes that
$0$ mag
$\equiv$
$1\,\myelectrons$ (electron)
$\equiv$
$1\,\myphotons$ (photon)
for a Point Response Function volume of one $(\myvolume=1)$.}
($251\!\leq\!\myetrue\!\leq\!10^6~\myelectrons$),
and
a flat background was assumed with a value of $\myb=100$ $\myelectrons$.
Photon and readout noise was simulated, respectively,
using Poisson and Gaussian random noise generators and the resulting
observed background sky measurement error was
$\mysigmab=0.18$ $\myelectrons$.
The median effective-background area of the PRF of these observations was
$\mybeta=21.44$ $\mypx^2$.
All the simulated observations were analyzed with $\mympd$ using
an {\em{analytical}} Gaussian PSF with $\myfwhm\equiv3.0$ $\mypx$.

\begin{center}
\begin{figure}
\includegraphics[scale=0.70,trim= 10 175 360 110]{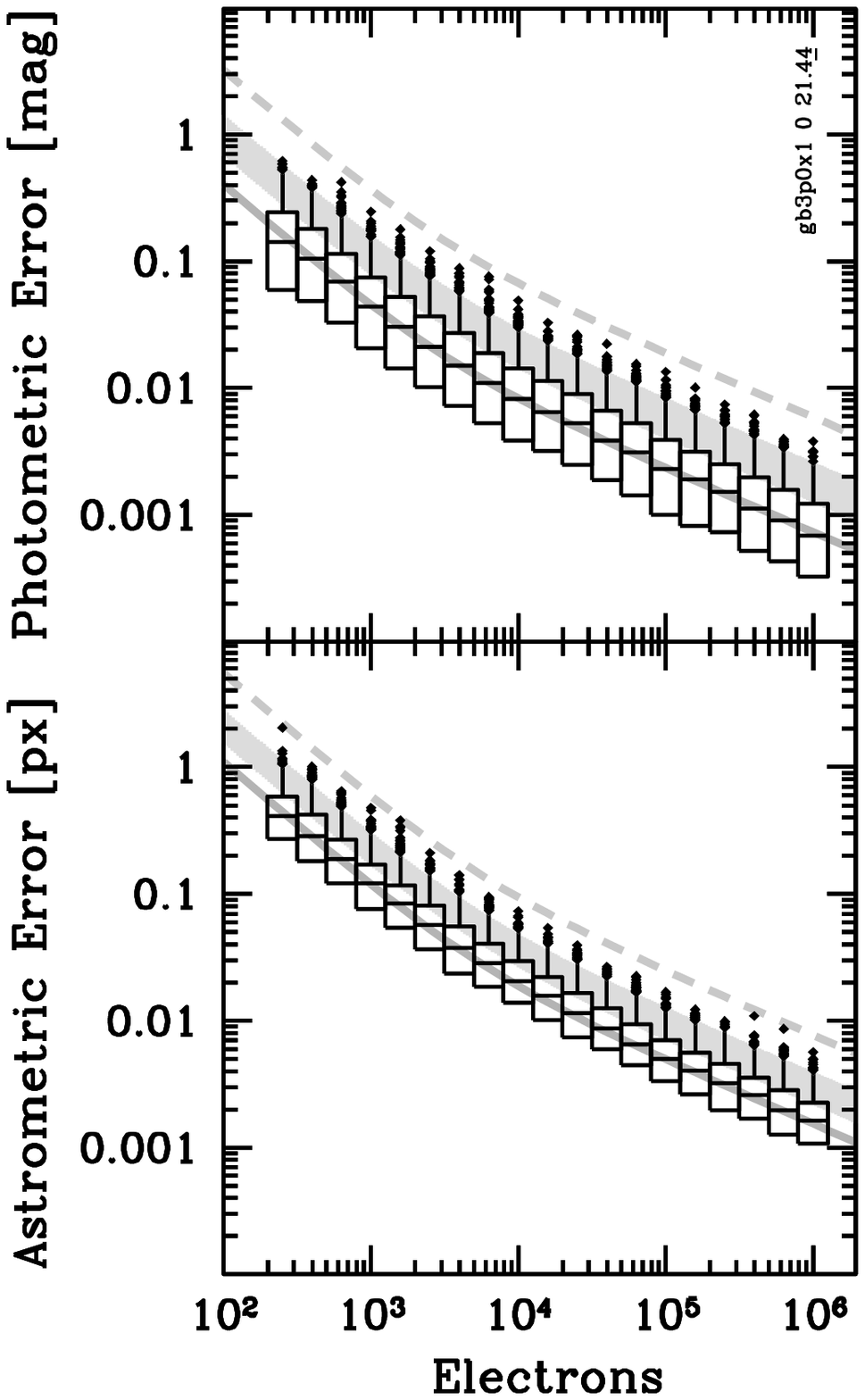}
\caption{The absolute photometric errors ({\em{top}})
and total astrometric errors ({\em{bottom}})
of $20,000$ simulated CCD stellar observations analyzed with $\mympd$
using an oversampled {\em{analytical}} Gaussian PSF with a
$\myfwhm$ of 3.0 $\mypx$ ($\mybeta\approx21.44$ $\mypx^2; \myvolume\equiv1)$.
\label{fig1}
}
\end{figure}

\end{center}
\begin{center}
\begin{figure}
\includegraphics[scale=0.70,trim= 10 175 360 290]{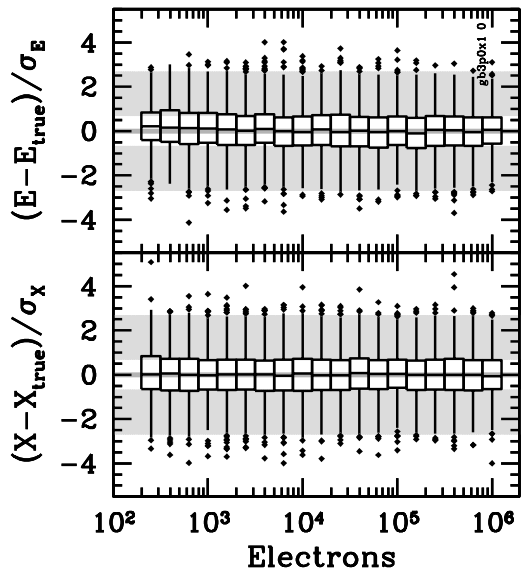}
\caption{Relative stellar intensity errors
({\em{top}})
and relative $\mysx$ position errors
({\em{bottom}})
of the data set used in Fig.~\protect\ref{fig1}.
\label{fig2}
}
\end{figure}
\end{center}

The binned absolute photometric errors are shown as black
box-and-whiskers plots (see Appendix \ref{appendix:a})
on the {\em{top}} panel of Fig.~\ref{fig1}.
The absolute photometric error of an observation
is the absolute value of the difference between the
measured (estimated) and true (actual) stellar magnitude:
$\Delta\mymag \equiv | \mymag - \mymagtrue |$.
The four {\em{grey}} limits seen on the {\em{top}} panel of Fig.~\ref{fig1}
are theoretical predictions (derived from \S 2.3.4)
for the median (50\% cumulative fraction: grey solid curve),
top hinge (75\%: bottom of the grey band),
top fence ($\sim$98.35\%: top of band),
and 5-$\sigma$ outlier ($\sim$99.99997\%: grey dashed curve) values.
If the rms photometric error is called $\sigma_{\mymag}$, then the
values of these theoretical limits are approximately equal to
$0.674\,\sigma_{\mymag}$,
$1.151\,\sigma_{\mymag}$,
$2.398\,\sigma_{\mymag}$,
and
$5.423\,\sigma_{\mymag}$, respectively.
If the photometric performance model is correct and
$\mympd$ has been coded correctly, then
(1) the observed median values (central bar in each box)
should intersect the theoretical median value,
(2) most of the top whiskers should be found inside the band,
and
(3) most of the outliers should be found
above the top of the band and all of the outliers should found
below the 5-$\sigma$ outlier limit.

Comparing the absolute photometric errors of the 20,000 simulated CCD
observations with the grey theoretical
limits, one sees that the photometric performance of the $\mympd$ code is
very well predicted by the model given in \S 2.3.4.

The binned total astrometric errors are shown as black
box-and-whiskers plots
on the {\em{bottom}} panel of Fig.~\ref{fig1}.
The total astrometric error of an observation
is the distance between the
measured (estimated) and true (actual) position of a star:
$\Delta\myr
\equiv
\sqrt{(\mysx\!-\!\mysx_{\rm{true}})^2\!+\!(\mysy - \mysy_{\rm{true}})^2}$.
The four {\em{grey}} limits seen on the {\em{bottom}} panel
of Fig.~\ref{fig1}
are theoretical predictions (derived from \S 2.4.4)
for the median (50\% cumulative fraction: grey solid curve),
top hinge (75\%: bottom of the grey band),
top fence ($\sim$98.97\%: top of band),
and 5-$\sigma$ outlier (99.99997\%: grey dashed curve) values.
The values of these theoretical limits are approximately equal to
$1.178\,\mysigmasx$,
$1.666\,\mysigmasx$,
$3.027\,\mysigmasx$,
and
$5.890\,\mysigmasx$,
where
$\mysigmasx$ is the
rms measurement error for the stellar $\mysx$ position.
If the astrometric performance model is correct and
$\mympd$ has been coded correctly, then
(1) the observed median values
should intersect the theoretical median value,
(2) most of the top whiskers should be found inside the band,
and
(3) most of the outliers should be found
above the top of the band and all of the outliers should found
below the 5-$\sigma$ outlier limit.

Comparing the total astrometric errors of the 20,000 simulated CCD
observations with the grey theoretical
limits, one sees that the astrometric performance of the $\mympd$ code is
very well predicted by the model given in \S 2.4.4.

Figure~\ref{fig2} shows the {\em{relative}} stellar intensity errors
and the {\em{relative}} $\mysx$ position errors
of the 20,000 stars analyzed in Fig.~\ref{fig1}.
The relative stellar intensity error is the difference between the
measured (estimated) and true (actual) stellar intensity values
divided by the estimated stellar intensity error:
$\Delta\mye \equiv (\mye - \mye_{\rm{true}} ) / \mysigmae $.
The relative $\mysx$ position error is the difference between the
measured (estimated) and true (actual) stellar $\mysx$ position values
divided by the estimated $\mysx$ error:
$\Delta\mysx \equiv (\mysx - \mysx_{\rm{true}} ) / \mysigmasx $.
If $\mympd$ has been coded correctly, the relative error distributions
for the stellar parameters $\mye$, $\mysx$, and $\mysy$ should be
{\em{normally}} distributed.
The five {\em{grey}} limits seen on each panel
are theoretical predictions (based on the normal distribution)
for, from bottom to top, the
bottom fence ($\sim$0.35\% cumulative fraction:
bottom of the bottom grey band),
bottom hinge (25\%: top of bottom band),
median (50\%: grey solid line at zero),
top hinge (75\%: bottom of top band),
top fence ($\sim$99.65\%: top of top band) values.
If the relative errors for $\mye$ and $\mysx$ are indeed normally
distributed, then
(1) the observed median values
should be near zero,
(2) most of the whiskers should be found inside the bands,
and
(3) most of the outliers should be beyond the fence values.

Comparing the relative errors for
$\mye$ and $\mysx$ of the 20,000 simulated CCD
observations with the grey theoretical limits, one sees that
these errors are, as expected, normally distributed.

\begin{itemize}
\item
The $\mympd$ code works well with oversampled
analytical Gaussian PSFs and its performance can be very well predicted
with the photometric and astrometric models derived in \S 2.
\end{itemize}

\subsubsection{Discrete PSFs}

\begin{center}
\begin{figure}
\includegraphics[scale=0.70,trim= 10 175 360 110]{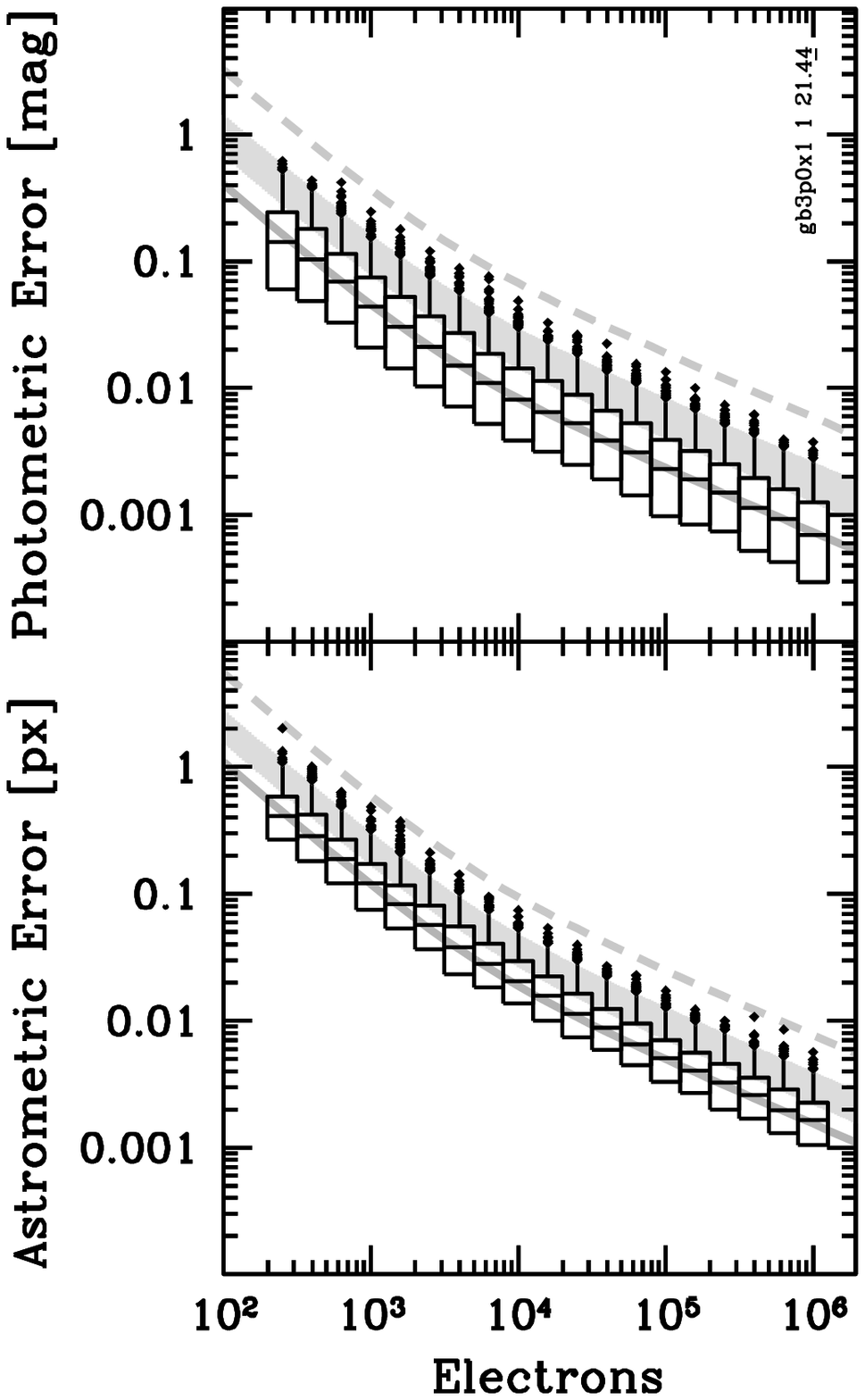}
\caption{The absolute photometric errors ({\em{top}})
and total astrometric errors ({\em{bottom}})
of $20,000$ simulated CCD stellar observations
used in Figs.~\protect\ref{fig1} and \protect\ref{fig2}
were analyzed with $\mympd$
using an oversampled {\em{discrete}} Gaussian PSF with a
$\myfwhm$ of 3.0 $\mypx$ ($\mybeta\approx21.44$ $\mypx^2; \myvolume\equiv1)$.
\label{fig3}
}
\end{figure}
\end{center}

\begin{center}
\begin{figure}
\includegraphics[scale=0.70,trim= 10 175 360 290]{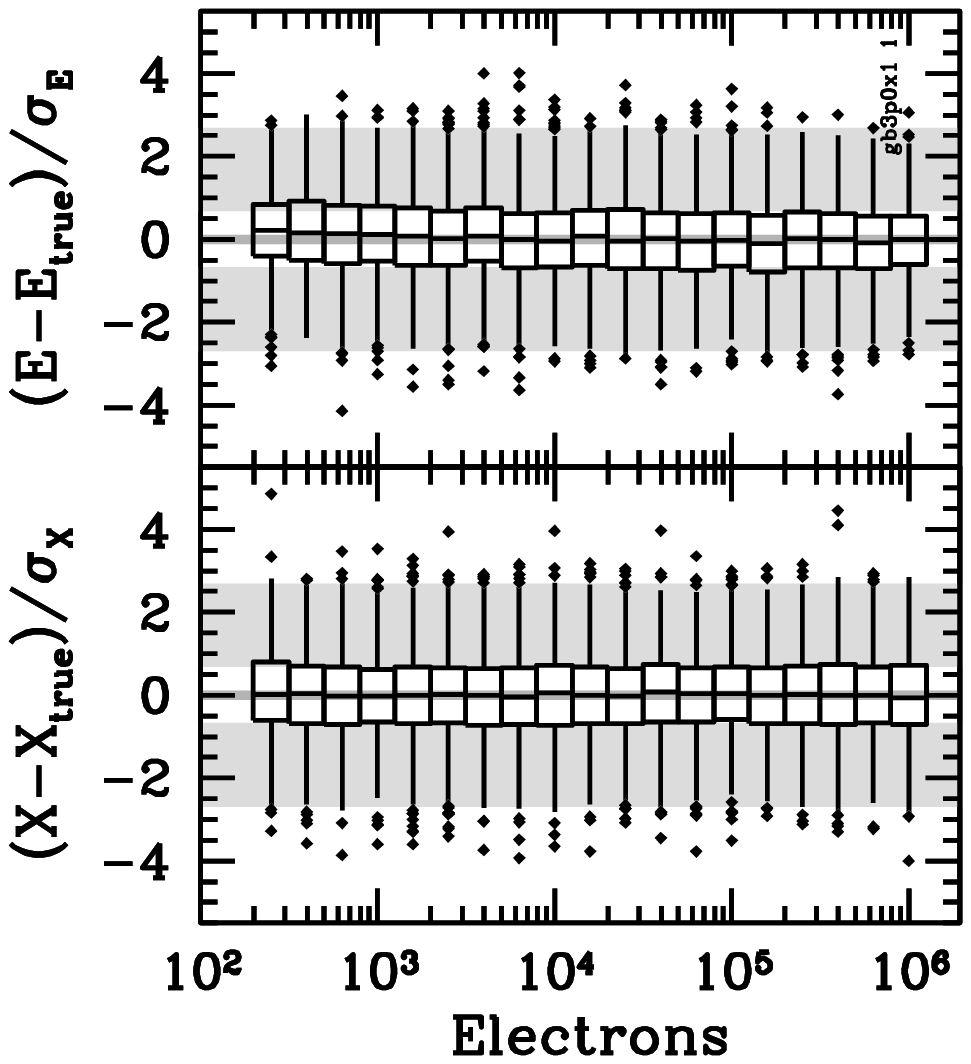}
\caption{Relative stellar intensity errors
({\em{top}})
and relative $\mysx$ position errors
({\em{bottom}})
of the data set used in Fig.~\protect\ref{fig3}.
\label{fig4}
}
\end{figure}
\end{center}

The 20,000 simulated CCD observations analyzed in
Figs.~\ref{fig1} and \ref{fig2}
were {\em{reanalyzed}} with $\mympd$
using an oversampled {\em{discrete}} Gaussian PSF with a
$\myfwhm$ of 3.0 $\mypx$.
Figure~\ref{fig3} shows the resultant
absolute photometric errors
and total astrometric errors.
Figure~\ref{fig4} shows the resultant
relative errors for
$\mye$ and $\mysx$.
Notice how similar
Figs.~\ref{fig1} and \ref{fig3}
and
Figs.~\ref{fig2} and \ref{fig4}
are to each other.

Despite the very different way the shape information of the PSF
was encoded (i.e., {\em{discrete}} versus {\em{analytical}} representations),
$\mympd$ produced nearly identical photometric and astrometric results.

\begin{center}
\begin{figure}
\includegraphics[scale=0.70,trim= 10 175 360 290]{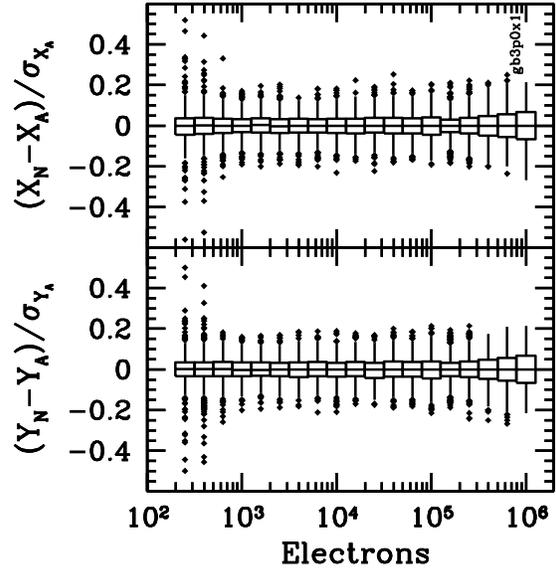}
\caption{Relative $\mysx$ and $\mysy$ position differences
({\em{top}} and {\em{bottom}}, respectively)
between the
numerical ({\em{subscript N}})
and
analytical ({\em{subscript A}})
results of the same 20,000 stars used in
Figs.~\protect\ref{fig1}--\protect\ref{fig4}.
\label{fig5}
}
\end{figure}
\end{center}

How similar are the measured stellar positions?
Figure~\ref{fig5} shows the {\em{relative}}
$\mysx$ and $\mysy$ position {\em{differences}}
between the previous
analytical
and
numerical
analyses with the $\mympd$ code.
The {\em{top}} panel shows the
difference between the
numerical $\mysx$ result
and the
analytical $\mysx$ result
divided by the estimated error of the analytical result.
Similarly,
the {\em{bottom}} panel shows the
difference between the
numerical $\mysy$ result
and the
analytical $\mysy$ result
divided by the estimated error of the analytical result.
The relative differences between the numerical and analytical methods
are {\em{not}} normally distributed --- observe how much smaller the values
on the ordinate of Fig.~\ref{fig5} are
compared to those of Figs.~\ref{fig2} and \ref{fig4}.
Figs.~\ref{fig2} and \ref{fig4} {\em{are}} normally distributed and the
source of the scatter is photon noise.
Figure~\ref{fig5} indicates that
the relative differences between the numerical and analytical methods for
astrometry are less than one-fifteenth the difference due to photon noise.
In other words,
{\em{the computational noise due to the chosen analysis method
(numerical versus analytical)
is insignificant when compared to the unavoidable photon noise
due to the random arrival of photons in any astronomical
CCD observation.}}

\begin{itemize}
\item
The $\mympd$ code works as well with oversampled discrete
Gaussian PSFs as it does with oversampled analytical Gaussian PSFs.
\end{itemize}

\begin{center}
\begin{figure}
\includegraphics[scale=0.70,trim= 10 175 360 110]{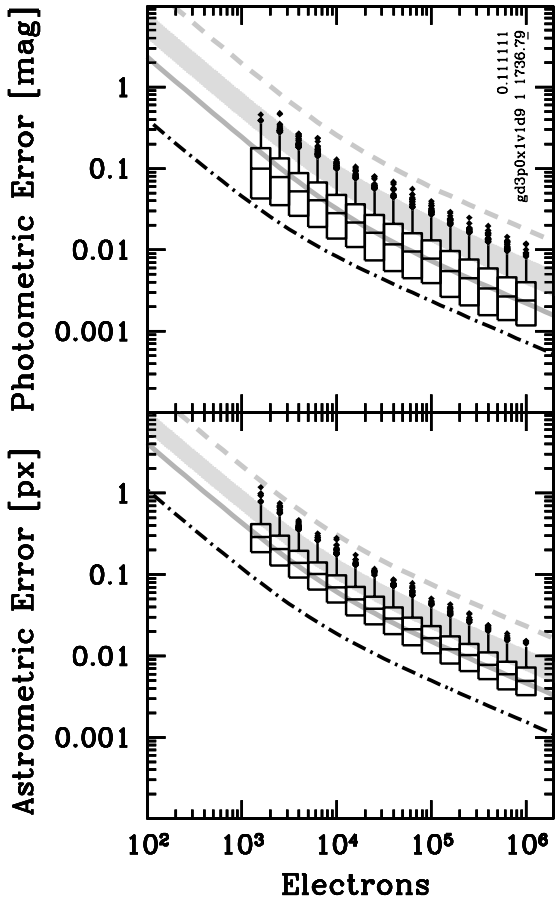}
\caption{The absolute photometric errors ({\em{top}})
and total astrometric errors ({\em{bottom}})
of $20,000$ simulated CCD stellar observations analyzed with $\mympd$
using a discrete Gaussian PSF with a
$\myfwhm$ of 3.0 $\mypx$ with an inefficient detector with $V\!=\!1/9$
$\,(\mybeta\approx1736.79$ $\mypx^2)$.
See the text for more details.
\label{fig6}
}
\end{figure}
\end{center}

\subsubsection{Inefficient Detectors}

While the volume, $\myvolume$, of the PRF was carefully
tracked throughout the derivation of the photometric
and astrometric performance models in \S 2,
all previous simulations have assumed a perfect detector
$(\myvolume\equiv1)$.
Let us now check to see if the effects of a PRF volume integral
that is less than one
has been correctly accounted for in the performance models of \S 2
by analyzing simulated observations imaged on a
very inefficient detector
$(\myvolume\ll1)$.

Twenty thousand oversampled CCD
stellar observations were simulated
assuming a very inefficient detector with $V=1/9$.
Stars were simulated using a discrete Gaussian PSF with a
$\myfwhm\equiv3$ $\mypx$ located near the center of 60$\times$60 pixels,
the input stellar intensities ranged from
$-8$ to $-15$ mag
($1585$ to $10^6$ $\myphotons$);
and
the {\em{observed}} background sky level was assumed to be a
constant value of $\myb=11.1111$ $\myelectrons$
($\myb_{\rm{true}}=100$ $\myphotons$, $\langle\myvolume\rangle=1/9$)
all other simulation parameters were the same as before.

All the simulated observations were analyzed with
$\mympd$ in the same way
as described for the numerical experiment shown in Fig.\ \ref{fig3}
--- except that the volume of the PRF was set to $\myvolume = 1/9$
in order to simulate the use of
an inefficient detector which converts only $\sim$11.1\%
of photons to electrons.

Figure~\ref{fig6} shows the
absolute photometric errors
and total astrometric errors of this numerical experiment.
The median effective-background area of PRF of these observations was
$\mybeta\approx1736.79$ $\mypx^2$ which is, as expected,
81 $(=\myvolume^{-2})$ times larger
than the median value reported in
Fig.\ \ref{fig3}.

Comparing the simulation results with the
grey theoretical limits, one sees that the
photometric and astrometric performance of the $\mympd$ code is
very well predicted by the theoretical performance models given in \S 2.

The black dash-dot curves in each panel of Fig.\ \ref{fig6}
shows the expected median response with a perfect detector;
these curves are the same as the
solid grey median curves found in
Fig.\ \protect\ref{fig3}.
The {\em{observed}} stellar intensities and observed background sky level
are nine times fainter than was seen in the numerical experiment
shown in Fig.\ \ref{fig3} and the
median photometric and astrometric errors
in Fig.\ \ref{fig6}
are, as expected, $\sim$3 $(=V^{-1/2})$ times larger
when the inefficient detector is used.

\begin{itemize}
\item
The $\mympd$ code and the theoretical performance models
work well with PRFs that have volumes of less than one.
\end{itemize}

\subsection{Undersampled Discrete PSFs}

Twenty thousand {\em{undersampled}} CCD
stellar observations were simulated using
an analytical Gaussian with a $\myfwhm\equiv1.5$ $\mypx$,
the other simulation parameters were the same as given in
\S 5.1.1.
The median effective-background area of PRF of these observations was
$\mybeta\approx6.12$ $\mypx^2$ $(\myvolume\equiv1)$.
All the simulated observations were analyzed with $\mympd$ using
a {\em{discrete}} Gaussian PSF with $\myfwhm\equiv1.5$ $\mypx$.

\begin{center}
\begin{figure}
\includegraphics[scale=0.70,trim= 10 175 360 110]{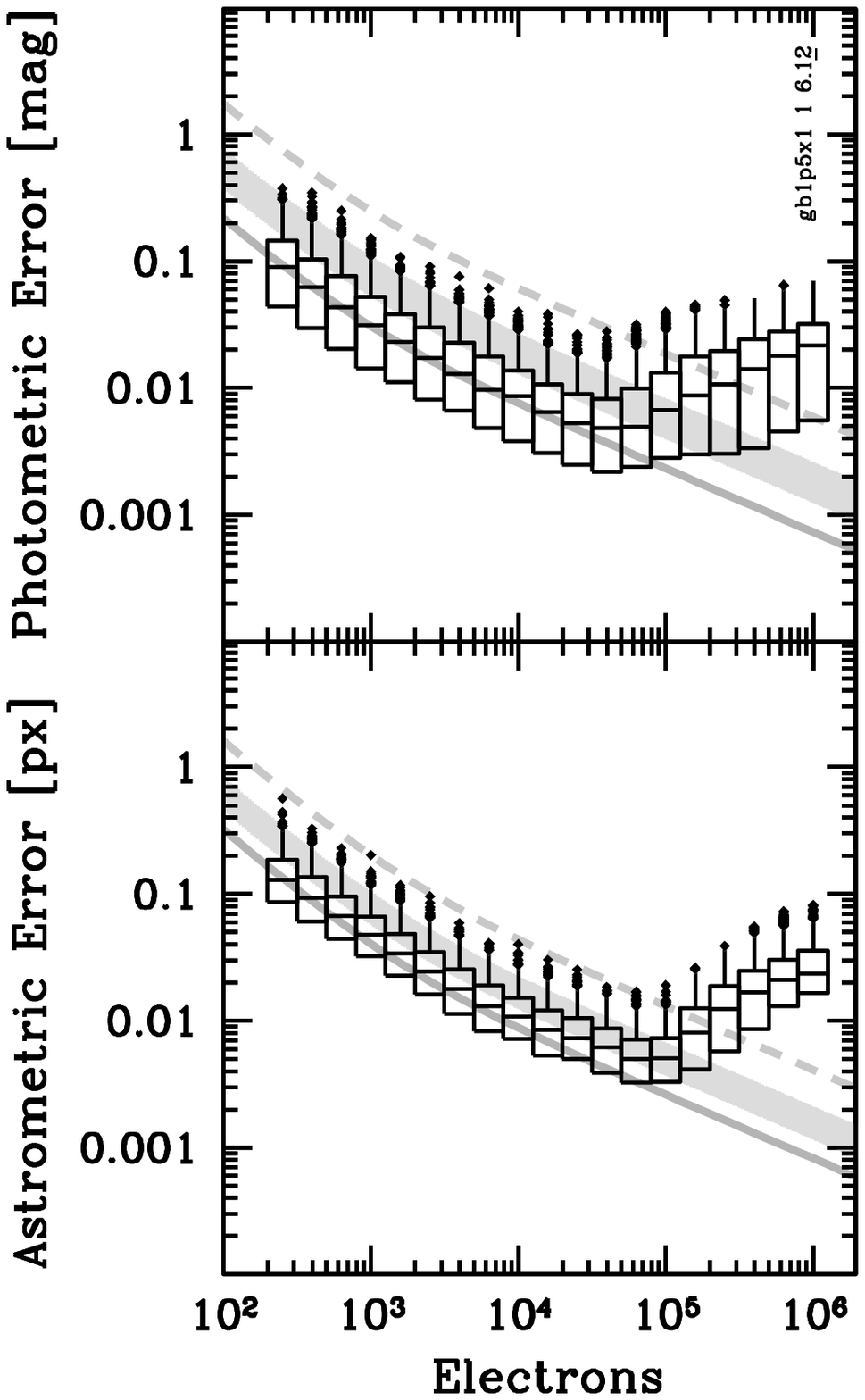}
\caption{The absolute photometric errors ({\em{top}})
and total astrometric errors ({\em{bottom}})
of $20,000$ simulated CCD stellar observations analyzed with $\mympd$
using an undersampled {\em{discrete}} Gaussian PSF with a
$\myfwhm$ of 1.5 $\mypx$
$(\mybeta\approx6.12$ $\mypx^2; \myvolume\!\equiv\!1 )$.
\label{fig7}
}
\end{figure}
\end{center}

Figure~\ref{fig7} shows the
absolute photometric errors
and total astrometric errors of this numerical experiment.
While the photometric and astrometric results for stars with
$\myetrue\!\la\!30,0000$ $\myelectrons$ are fine,
the results for stars brighter than that limit are seen to
quickly degrade in accuracy with the
brightest stars having median errors that are $\sim$40 times
worse than expected.

\begin{center}
\begin{figure}
\includegraphics[scale=0.40,trim= 0 125 0 150]{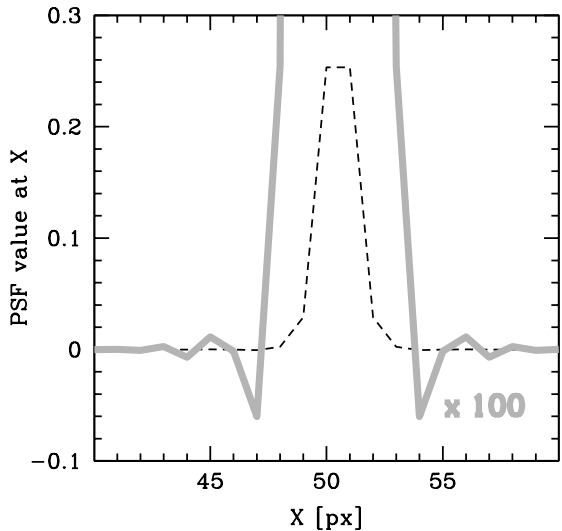}
\caption{A one pixel wide slice through a pixel-centered discrete Gaussian
PSF with $\myfwhm=1.5$ $\mypx$ that was shifted half of a
pixel in $X$ to the right using
equation (\protect\ref{eq:dampedsinc}).
The thick grey curve is the same PSF multiplied by a factor of 100.
\label{fig8}
}
\end{figure}
\end{center}

What starts going wrong at $\myetrue\approx30,000$ $\myelectrons$?
Figure~\ref{fig8} shows a one pixel wide slice through a pixel-centered
discrete Gaussian PSF with $\myfwhm=1.5$ $\mypx$ that was shifted half of a
pixel in $X$ to the right using damped sinc function given in
equation (\protect\ref{eq:dampedsinc}).
The dashed black curve looks fine, but when expanded by a factor of 100, one
sees that {\em{negative side lobes have been created}} due to the fact that
the Nyquist-Shannon Sampling Theorem has been violated.
Doing a sinc interpolation (damped or otherwise) on undersampled data
is never a good idea -- the ``ringing'' seen in Fig.~\ref{fig8} is
a classic signature of an edge that is too sharp to be adequately
expressed with the limited spatial information contained in an
undersampled observation.
The biggest negative side lobe of the shifted PSF
has a value of about $-$0.0006. Although that may seem to be a small
value compared to the total volume integral of one, it is actually
quite disastrous because {\em{negative PSF values have no
physical meaning}}.

It is now clear what has gone wrong for stars with
$\myetrue \ga 30,000$ $\myelectrons$.
At stellar intensity values greater than 17,000 electrons, the
intensity-scaled undersampled PSF models can have
negative side lobes that are larger
than the rms observed background sky noise level
$(\,|$$-$0.0006$| * 17,000\,\myelectrons
= 10.2\,\myelectrons
> 10\,\myelectrons
\approx
\sqrt{B})$.
At stellar intensity values greater than 167,000 electrons, the
observational models have physically nonsensical {\em{negative sky values}}.

\begin{itemize}
\item
Aliasing (ringing) effects will
generally only be seen with bright stars since a large number of
photons are required in order to adequately sample the higher spatial
frequencies of a PSF.
\item
Fitting undersampled observations of
bright stars with undersampled PSFs
results in poor photometry and astrometry.
\end{itemize}

\subsection{Supersampled Discrete PSFs}

\begin{center}
\begin{figure}
\includegraphics[scale=0.70,trim= 10 175 360 110]{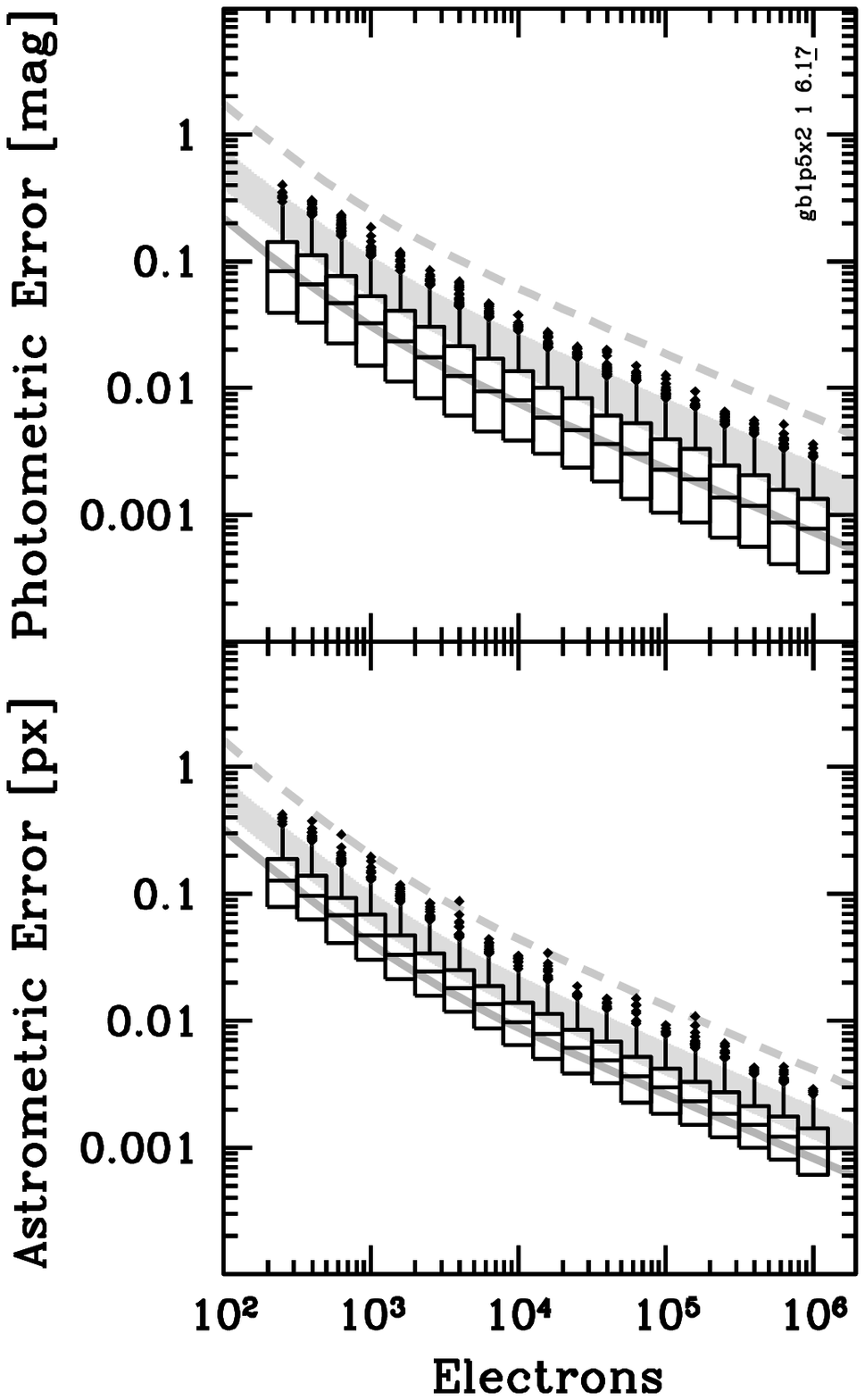}
\caption{The absolute photometric errors ({\em{top}})
and total astrometric errors ({\em{bottom}})
of $20,000$ simulated CCD stellar observations analyzed with $\mympd$
using a {\em{2$\times$2 supersampled}} discrete Gaussian PSF with a
$\myfwhm$ of 1.5 $\mypx$
$(\mybeta\approx6.17$ $\mypx^2; \myvolume\!\equiv\!1 )$.
\label{fig9}
}
\end{figure}
\end{center}

\begin{center}
\begin{figure}
\includegraphics[scale=0.70,trim= 10 175 360 290]{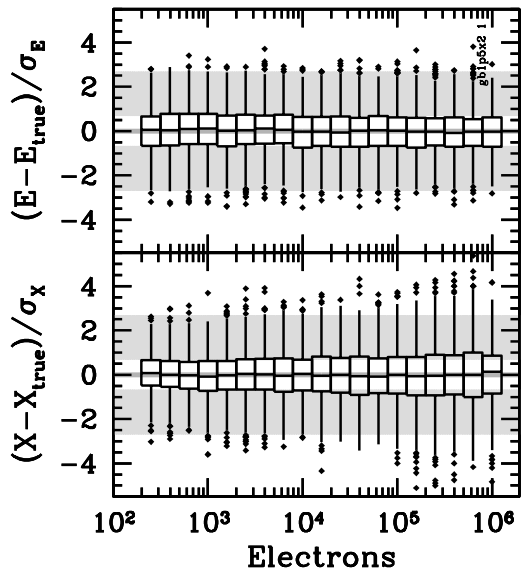}
\caption{Relative stellar intensity errors
({\em{top}})
and relative $\mysx$ position errors
({\em{bottom}})
of the data set used in Fig.~\protect\ref{fig9}.
\label{fig10}
}
\end{figure}
\end{center}

A supersampled PSF is a PSF with pixels that have greater spatial resolution
(higher spatial frequencies) than the actual pixels in the observational data.
For example, a 2$\times$2 supersampled PSF uses 4 pixels to describe every
physical pixel of the CCD observation; each supersampled pixel has twice
the spatial resolution of the actual pixels in the observation.

Twenty thousand {\em{undersampled}} CCD
stellar observations were simulated using
an analytical Gaussian with a $\myfwhm\equiv1.5$ $\mypx$;
the other simulation parameters were the same as
before.
All the simulated observations were analyzed with $\mympd$ using
a 2$\times$2 {\em{supersampled}} discrete Gaussian PSF with
$\myfwhm\equiv1.5$ $\mypx$ ($\mybeta\approx6.17$ $\mypx^2$;
$\myvolume\equiv1$).

Figure~\ref{fig9} shows the
absolute photometric errors
and total astrometric errors of this numerical experiment.
By providing $\mympd$ with extra information, in the form of a
supersampled PSF, the Nyquist-Shannon Sampling Theorem was no longer
violated and excellent photometry and astrometry was done with
this undersampled data set.

Figure~\ref{fig10} shows the
relative errors for
$\mye$ and $\mysx$.
The relative stellar intensity errors are normally distributed.
However, the relative $\mysx$ position errors
are almost, but not quite, normally distributed.
The $\mympd$ code is accurately measuring the stellar positions
(i.e., the median difference, $\mysx - \mysx_{\rm{true}}$, values are zero)
but the rms position error estimates
($\mysigmasx$) are slightly {\em{underestimated}}
(the top and bottom whiskers for $\myetrue \ga 10,000$ $\myelectrons$ are seen
to extend beyond the grey bands).
The same effect is seen with $\mysy$.
Using a higher-resolution supersampled PSF (3$\times$3, 4$\times$4, $\ldots$)
does not eliminate the small underestimation by $\mympd$ of position errors.
{\em{The position errors estimated
by $\mympd$ are close to the photonic limit,
but the actual errors
--- for undersampled observations
--- are close to the astrometric CRLB with square CCD pixels}}
\citep{winick:1986}.

\begin{itemize}
\item
Accurate and precise CCD stellar photometry and astrometry may be
obtained with undersampled CCD observations if supersampled PSFs are
used during the PSF-fitting process.
\end{itemize}

\subsection{Critically-Sampled Discrete PSFs}

Let us now investigate what happens when critically-sampled data is
fit with a critically-sampled PSF.

Twenty thousand {\em{critically-sampled}} CCD
stellar observations were simulated using
an analytical Gaussian with a $\myfwhm\equiv2.35482$ $\mypx$;
the other simulation parameters were the same as
before.
All the simulated observations were analyzed with $\mympd$ using
a {\em{critically-sampled}} discrete Gaussian PSF with
$\myfwhm\equiv2.35482$ $\mypx$ ($\mybeta\approx13.62$ $\mypx^2$;
$\myvolume\equiv1$).

\begin{center}
\begin{figure}
\includegraphics[scale=0.70,trim= 10 175 360 110]{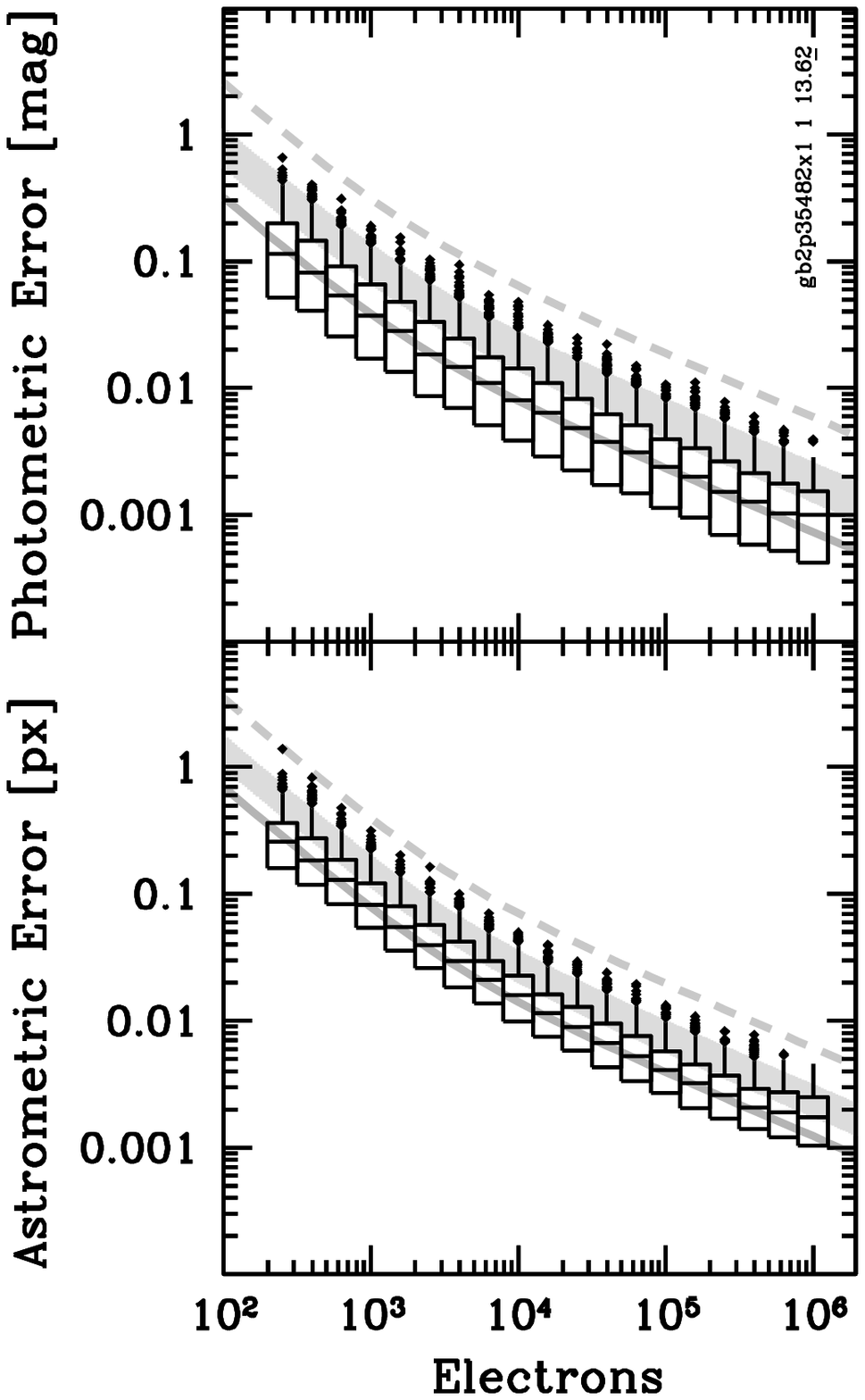}
\caption{The absolute photometric errors ({\em{top}})
and total astrometric errors ({\em{bottom}})
of $20,000$ simulated CCD stellar observations analyzed with $\mympd$
using critically-sampled discrete Gaussian PSF with a
$\myfwhm$ of 2.35482 $\mypx$
$(\mybeta\approx13.62$ $\mypx^2; \myvolume\!\equiv\!1 )$.
\label{fig11}
}
\end{figure}
\end{center}

\begin{center}
\begin{figure}
\includegraphics[scale=0.70,trim= 10 175 360 290]{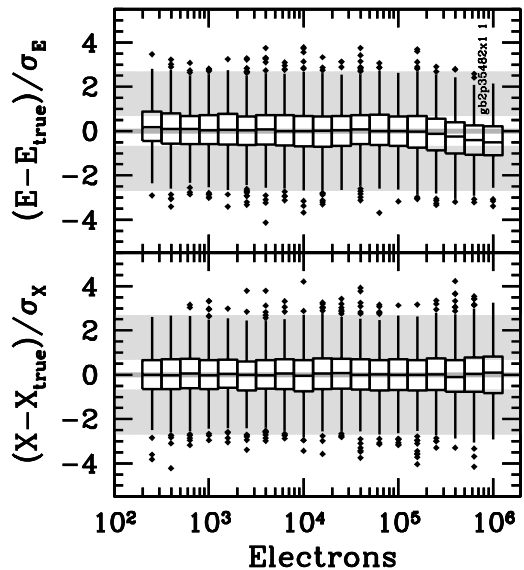}
\caption{Relative stellar intensity errors
({\em{top}})
and relative $\mysx$ position errors
({\em{bottom}})
of the data set used in Fig.~\protect\ref{fig11}.
\label{fig12}
}
\end{figure}
\end{center}

Figure~\ref{fig11} shows the
absolute photometric errors
and total astrometric errors of this numerical experiment.
Figure~\ref{fig12} shows the
relative errors for
$\mye$ and $\mysx$.
Looking carefully at Figs.~\ref{fig11} and \ref{fig12}, one sees that the
photometric and astrometric performance is well matched to the theoretical
expectations except for the brightest three bins
($\myetrue \ga 316,000$ $\myelectrons$).

Twenty thousand {\em{critically-sampled}} CCD
stellar observations were simulated using
an analytical Gaussian with a $\myfwhm\equiv2.35482$ $\mypx$;
the other simulation parameters were the same as
before.
All the simulated observations were analyzed with $\mympd$ using
a 2$\times$2 {\em{supersampled}} discrete Gaussian PSF with
$\myfwhm\equiv2.35482$ $\mypx$ ($\mybeta\approx13.62$ $\mypx^2$;
$\myvolume\equiv1$).

\begin{center}
\begin{figure}
\includegraphics[scale=0.70,trim= 10 175 360 110]{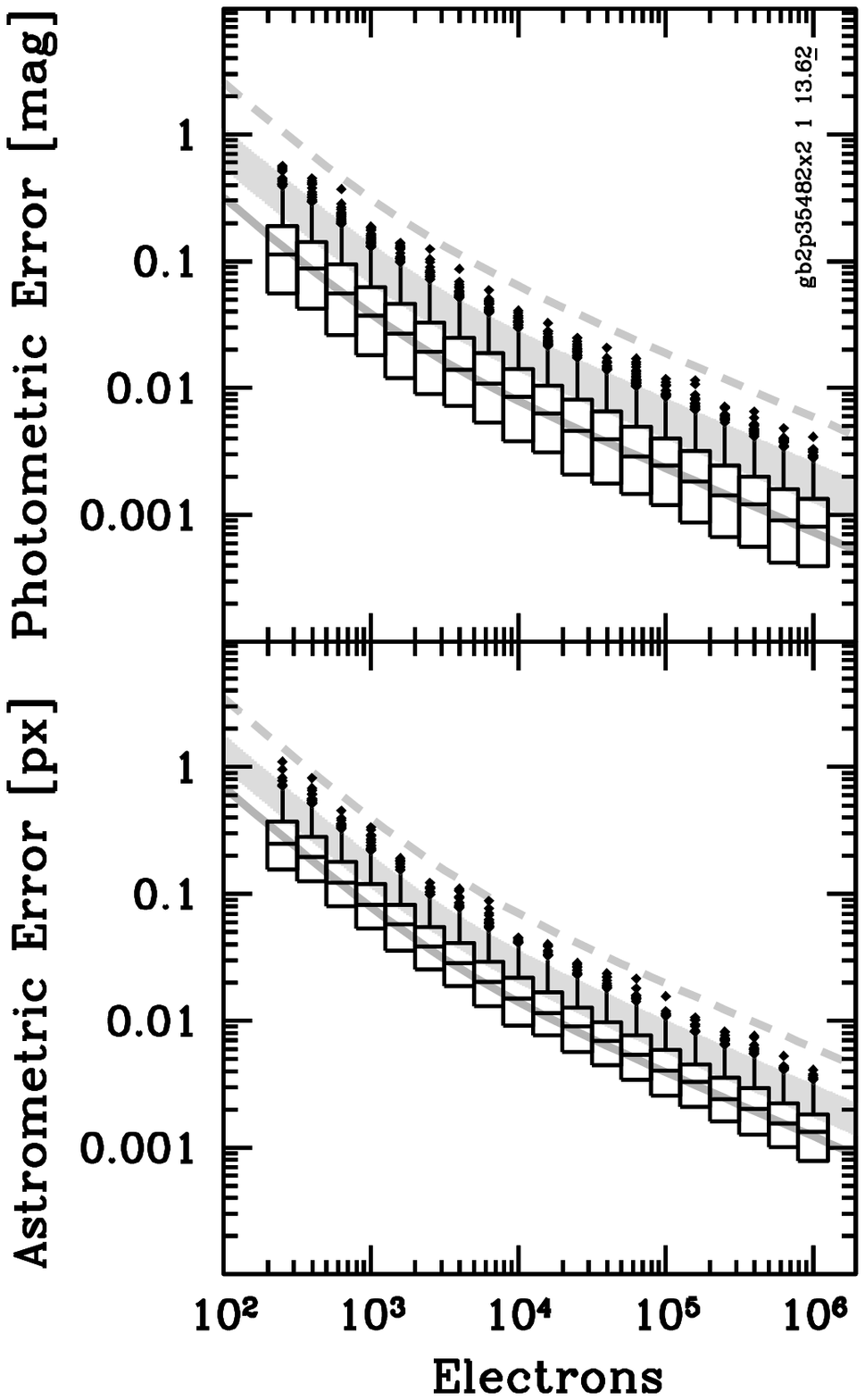}
\caption{The absolute photometric errors ({\em{top}})
and total astrometric errors ({\em{bottom}})
of $20,000$ simulated CCD stellar observations analyzed with $\mympd$
using a {\em{2$\times$2 supersampled}} discrete Gaussian PSF with a
$\myfwhm$ of 2.35482 $\mypx$
$(\mybeta\approx13.62$ $\mypx^2; \myvolume\!\equiv\!1 )$.
\label{fig13}
}
\end{figure}
\end{center}

\begin{center}
\begin{figure}
\includegraphics[scale=0.70,trim= 10 175 360 290]{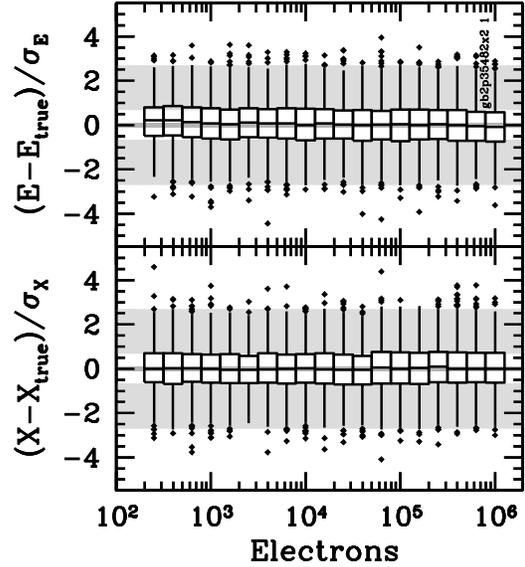}
\caption{Relative stellar intensity errors
({\em{top}})
and relative $\mysx$ position errors
({\em{bottom}})
of the data set used in Fig.~\protect\ref{fig13}.
\label{fig14}
}
\end{figure}
\end{center}

Figure~\ref{fig13} shows the
absolute photometric errors
and total astrometric errors of this numerical experiment.
Figure~\ref{fig14} shows the
relative errors for
$\mye$ and $\mysx$.
The
photometric and astrometric performance is well matched to the theoretical
expectations for all stellar intensities.

Comparing Fig.~\ref{fig7} with Fig.~\ref{fig9} and
Fig.~\ref{fig11} with Fig.~\ref{fig13}, one sees that
one can obtain excellent stellar photometry and astrometry
with the MATPHOT algorithm for all stellar intensities -- even if the
observational data is undersampled -- as long as
the discrete PSFs used to do the model fitting are sampled finely enough
to have sufficient spatial frequency coverage such that
the Nyquist-Shannon Sampling Theorem is not violated.

Comparing Fig.~\ref{fig3} with Fig.~\ref{fig11}, one sees that
the breakpoint for the MATPHOT algorithm
between undersampled and oversampled data
is $13.62<\mybeta\leq21.44$ $\mypx^2$ or,
in terms of a Gaussian Full-Width-at Half Maximum,
$2.35482\la\myfwhm\leq3$ $\mypx$.

\begin{itemize}
\item
If a discrete PSF is close to being critically sampled,
then one should use a supersampled discrete PSF
which is oversampled in terms of supersampled pixels $(\myspx)$.
In other words,
{\em{if the equivalent-background area is less than 21 square pixels
($\mybeta < 21$ $\mypx^2$; Gaussians: $\myfwhm < 3.0$ $\mypx$),
then one should use a supersampled discrete PSF which has
an equivalent-background area of
at least 21 square supersampled pixels
($\mybeta \geq 21$ $\myspx^2$; Gaussians: $\myfwhm \geq 3.0$ $\myspx$).
}}
\end{itemize}

\subsection{Ugly Discrete PSFs}

Let us now investigate the photometric and astrometric performance
of the MATPHOT algorithm with an ugly (realistic) space-based PSF.

\begin{center}
\begin{figure}
\includegraphics[scale=0.35,trim= 0 110 0 110]{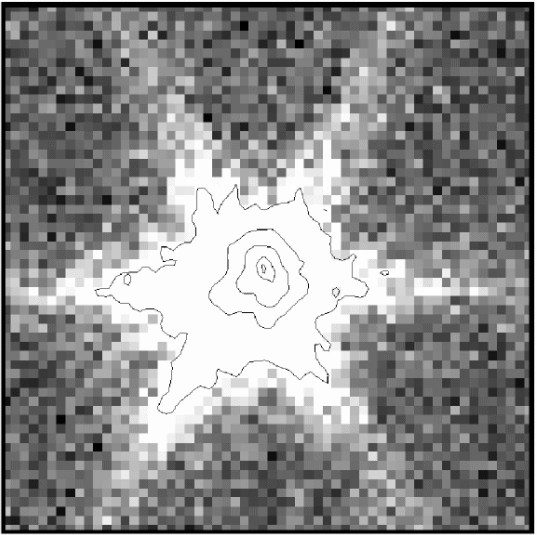}
\caption{A simulated $V$-band {\em{Next Generation Space Telescope}}
image based on a 2$\times$2 supersampled
PSF model for a 8-meter TRW-concept 1.5-micron
diffraction-limited primary mirror with 1/13 rms wave errors.
Contour levels of 90\%, 50\%, 10\%, 1\%, and 0.1\% of the peak intensity
are shown with {\em{black curves}}.
The pixel scale is 0.0128 arcsec $\mypx^{\,-1}$.
This image uses a linear stretch with a pixel intensity mapping of
{\em{black}} for $\la\,70$ $\myelectrons$
and {\em{white}} for $\ga\,150$ $\myelectrons$.
\label{fig15}
}
\end{figure}
\end{center}

Figure~\ref{fig15} shows a simulated
{\sl{Next Generation Space Telescope}} ({\sl{NGST}})
$V$-band CCD stellar observation.
This simulated observation used a
2$\times$2 supersampled PSF which was based on a
8-meter
TRW-concept
$1.5\,\mu$ diffraction-limited
primary mirror
with 1/13 wave rms errors at $1.5\,\mu$;
the original version of the PSF was kindly provided by John Krist (STScI).
The six-sided ``snowflake'' pattern seen in Fig.~\ref{fig15} is
mainly due to fact that the primary mirror is composed of
segmented hexagonal-shaped mirrors.
Observers will note that this PSF is very similar
to optical PSFs seen with
the 10-m telescopes at the W. M. Keck Observatory.
The 6.5-m {\sl{James Webb Space Telescope}} ({\sl{JWST}})
is likely to have similar looking near-infrared PSFs
once it achieves first light in $\sim$2011.

The {\sl{NGST}} PSF is so complicated that it is unlikely that it could
be represented adequately with a continuous analytical mathematical function.
Space-based observations frequently have high spatial frequencies
which make them ideal candidates for
photometric and astrometric analysis using discrete PSFs.

Twenty thousand CCD stellar observations were simulated using
the simulated $V$-band {\sl{NGST}}
2$\times$2 supersampled PSF described above;
the other simulation parameters were the same as before.
All the simulated observations were analyzed with $\mympd$ with the
PSF used to create the simulated observations.

\begin{center}
\begin{figure}
\includegraphics[scale=0.70,trim= 10 175 360 110]{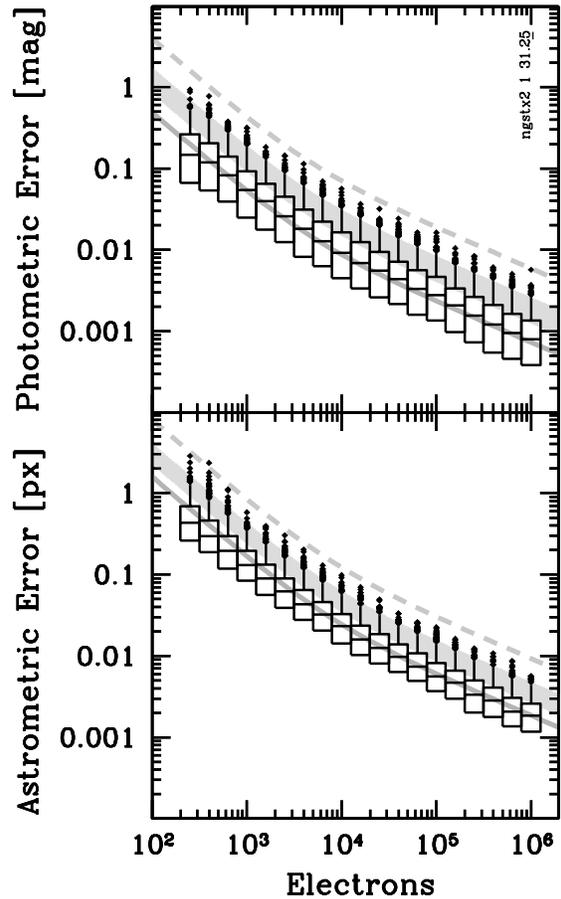}
\caption{The absolute photometric errors ({\em{top}})
and total astrometric errors ({\em{bottom}})
of $20,000$ simulated CCD stellar observations analyzed with $\mympd$
using the 2$\times$2 supersampled {\em{NGST}}
PSF described in Fig.~\protect\ref{fig15}
$(\mybeta\approx31.25$ $\mypx^2; \myvolume\!\equiv\!1 )$.
\label{fig16}
}
\end{figure}
\end{center}
\begin{center}
\begin{figure}
\includegraphics[scale=0.70,trim= 10 175 360 290]{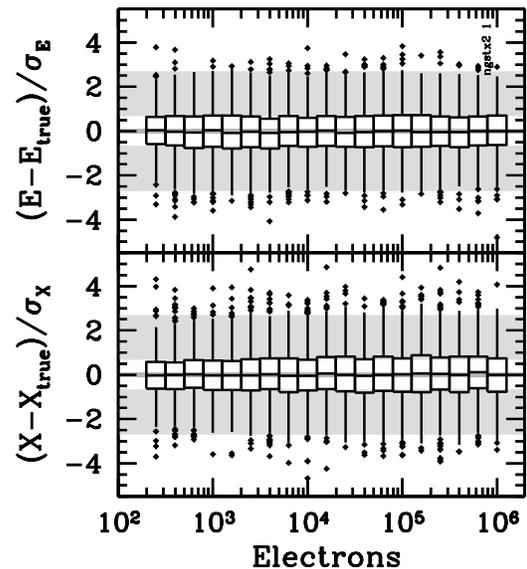}
\caption{Relative stellar intensity errors
({\em{top}})
and relative $\mysx$ position errors
({\em{bottom}})
of the data set used in Fig.~\protect\ref{fig16}.
\label{fig17}
}
\end{figure}
\end{center}

Figure~\ref{fig16} shows the
absolute photometric errors
and total astrometric errors of this numerical experiment.
Figure~\ref{fig17} shows the
relative errors for
$\mye$ and $\mysx$.
The
photometric and astrometric performance is well matched to the theoretical
expectations for all stellar intensities.

Although only Gaussian PSFs were used in previous numerical experiments,
the excellent fit seen in the top panel of Fig.~\ref{fig16} between
the theoretical photometric performance model (\S 2.3.4)
and actual $\mympd$ measurements using such an ugly discrete PSF
is not surprising once one remembers that
the theoretical photometric performance model was derived from an abstract
Point Response Function.

The development of the theoretical astrometric performance model, however,
required differentiation of the Point Response Function
which I assumed to be an oversampled analytical Gaussian function.
The analytical Gaussian bright star astrometric limit
was transformed to the general form
by assuming that the Gaussian-specific
$\myssigma^2$ term could be
replaced with the more general $\mycssl^2$ term, which, by definition,
can be computed for any PRF.
The same assumption was then used to
derive the general faint star astrometric limit.
The excellent fit seen in the bottom panel of Fig.~\ref{fig16}
indicates that this bold assumption is not only useful but practical.
Many numerical experiments with very ugly discrete PSFs
have shown that the
theoretical astrometric performance model of \S 2.4.4
works well with ugly discrete Point Response Functions.

\begin{center}
\begin{figure}
\includegraphics[scale=0.40,trim= 10 175 0 100]{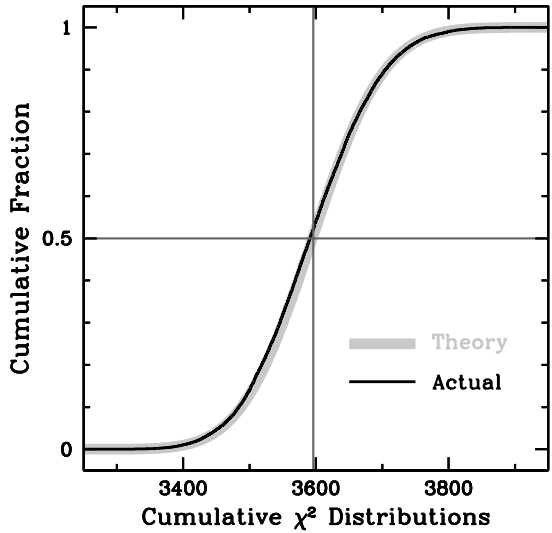}
\vspace*{1em}
\caption{A comparison between the cumulative $\chi^2$ distribution
for 3596
degrees of freedom ({\em{thick curve}})
and the measured $\chi^2$ value ({\em{thin curve}}),
of the data set used in Fig.~\protect\ref{fig16},
reported by the $\mympd$ implementation of the MATPHOT algorithm.
\label{fig18}
}
\end{figure}
\end{center}

If the MATPHOT algorithm is optimally extracting photometric and astrometric
information from a stellar observation,
{\em{and}} $\mympd$ has been correctly coded,
{\em{and}} the CCD observation has been properly calibrated,
{\em{and}} the PRF used in the observational model is correct,
{\em{and}} accurate estimates of the measurements errors for each pixel have
been made,
{\em{then}} one expects that the $\chi^2$
goodness-of-fit value reported by $\mympd$ to be distributed
as a $\chi^2$ distribution with the number of degrees of freedom equal
to the difference between the number of pixels in the observation and
the number of free parameters.
Figure \ref{fig18} shows that this prediction
about the precision and accuracy of the MATPHOT algorithm
has been verified:
the cumulative distribution of the $\chi^2$ reported by $\mympd$
({\em{thin black curve}}) is seen to lie on top of
the cumulative $\chi^2$ distribution of
for 3596
[$=$ $60^2$ pixels $-$ 4 free parameters
$(\mye$, $\mysx$, $\mysy$, and $\myb)$\,]
degrees of freedom ({\em{thick grey curve}}).

\begin{itemize}
\item
The $\mympd$ code works well with ugly discrete PSFs and its performance
can be well predicted using the general
theoretical photometric and astrometric performance models given
in \S 2.
\item
The $\chi^2$ goodness-of-fit value reported by $\mympd$ is a
statistically reliable measure of the quality of a
photometric and astrometric reduction of a stellar observation
obtained with the MATPHOT algorithm using ugly discrete PSFs.
\end{itemize}

\subsection{Ugly Detectors}

Let us now investigate the photometric and astrometric performance
of the MATPHOT algorithm with an ugly PSF and an ugly detector.

Suppose one has a detector where every
pixel has been divided up into 16 square regions called ``subpixels''.
Let us call the first row and first column of subpixels
``gate structures'' which are optically inactive with 0\% QE.
The remaining 9 subpixels are the optically active part of the pixel with
a 100\% QE.  By definition, such a pixel would have a very large
intrapixel QE variation with only 56.25\% of the total pixel area being
capable of converting photons to electrons.

A few extra lines of code were added to the
$\mympd$ program to simulate the image formation process with
such an ugly detector.
The new version of $\mympd$ is called $\mympdx$ and was designed
specifically for use with 4$\times$4 supersampled PSFs.

Ten thousand CCD stellar observations
of $-13$ mag stars
($\sim$$2.512^{13}$ $\myphotons$)
were simulated and analyzed with $\mympdx$ using
a 4$\times$4 supersampled version of the
simulated $V$-band {\sl{NGST}} PSF described above.
The {\em{observed}} background level was assumed to be a constant value of
$\myb=56.25$ $\myelectrons$
($\myb_{\rm{true}}=100$ $\myphotons$, $\langle\myvolume\rangle=0.5625$)
and
all other simulation parameters were the same as before.
The measured PRF volume of these simulated observations
was $\myvolume = 0.5616 \pm 0.0185$
which is consistent with the expected value of 0.5625 from the
physical structure of a single pixel.
The median and semiquartile range of the
effective-background area $(\mybeta)$
of these observations was, respectively,
28.10 and 4.82 $\mypx^2\,$.
The median critical-sampling scale length of these observations
was $\mycssl \approx 0.8398$ $\mypx$ --- indicating
that these observations were {\em{undersampled}}, as expected.

\begin{center}
\begin{figure}
\includegraphics[scale=0.35,bb= 0 100 574 634]{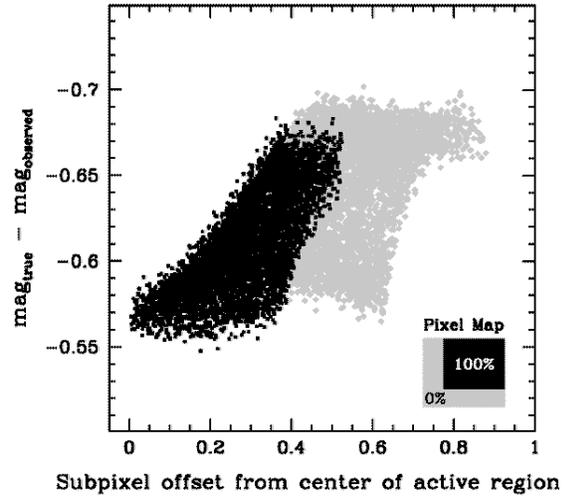}
\caption{The measured electron loss of the 10,000
simulated CCD observations of $-13$ mag stars
analyzed with $\mympdx$
using a 4$\times$4 supersampled version of the {\em{NGST}} PSF.
The electron loss is plotted as a function of the absolute value of the
distance from the center of a star and the center of the
active region of the central pixel of the stellar image.
\label{fig19}
}
\end{figure}
\end{center}

The optically
inactive gate structures of the pixel cause the observed number of
electrons in each stellar image to be significantly less than the number of
photons which fell on the detector.
{\em{The total amount of loss was dependent on where the center of the star
fell within the central pixel of the stellar image.}}
Figure\ \ref{fig19} shows that
stars centered in the middle of the active area of a
pixel suffered a $\sim$40\% loss
($\Delta m \approx 0.56$ mag)
while those centered on gate structures {\em{(grey points)}}
lost up to 47\%
($\Delta m \approx 0.69$ mag).

Although this numerical experiment may seem to be very artificial,
large intrapixel sensitivity variations can be found in cameras
currently installed on the {\sl{Hubble Space Telescope}}.
\citet{lauer99} reported peak-to-peak variations of
0.39 mag at the $J$ band (F110W)
and
0.22 mag at the $H$ band (F160W)
of the NIC3 camera
of the {\sl{HST}} NICMOS instrument.
The peak-to-peak variation of $\sim$0.2 mag
at F160W with NIC 3 was independently confirmed by
\citet{hookfruchter00}.

The mean {\em{observed}} stellar magnitude for these
$-13$ mag stars was $-12.3728\pm0.0359$ mag.  The photometric performance model
predicts an rms measurement error of $0.0036$ mag for these bright stars.
With an average loss of 44\% and an rms measurement error that is
{\em{ten times larger}} than expected from photon statistics,
the observed stellar magnitudes were neither precise or accurate.

\begin{center}
\begin{figure}
\includegraphics[scale=0.41,bb= 0 171 574 634]{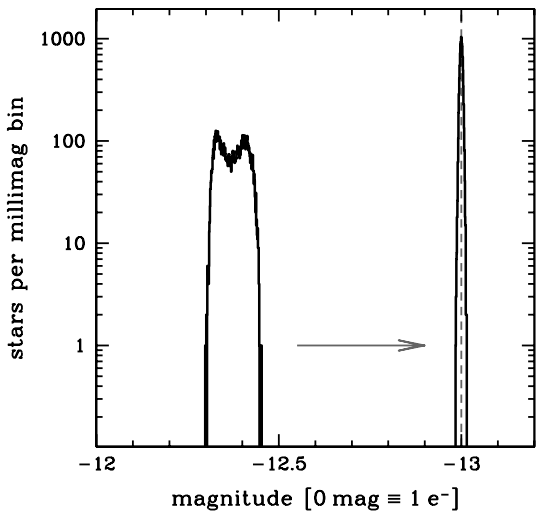}
\caption{The observed ({\em{left}}) and the measured ({\em{right}})
stellar magnitude distributions of the
10,000 $-13$ mag stars described in Fig.\ \protect\ref{fig19}.
\label{fig20}
}
\end{figure}
\end{center}

Figure \ref{fig20} shows that $\mympdx$ was able to do an excellent job
in recovering the true stellar magnitude of the 10,000 $-13$ mag stars
 --- despite
being presented with a worst-case scenario of undersampled observations
with an ugly PSF imaged on an ugly detector with a very large
intrapixel QE variation.

The mean {\em{measured}} stellar magnitude reported by $\mympdx$ was
$-12.9998  \pm 0.0039$ mag and the
mean rms error estimated by $\mympdx$ was $0.00384\pm0.00006$ mag.
The photometric performance of $\mympdx$ is fully consistent with
theoretical expectations --- which were
derived for an ideal detector with no intrapixel QE variation.

Twenty thousand CCD stellar observations
were simulated and analyzed with $\mympdx$ using
the same 4$\times$4 supersampled version of the
simulated $V$-band {\sl{NGST}} PSF.
The input stellar intensities ranged from
$-6$ to $-15$ mag
($251$ to $10^6$ $\myphotons$).
The observed background level was assumed to be a constant value of
$\myb=56.25$ $\myelectrons$
($\myb_{\rm{true}}=100$ $\myphotons$, $\langle\myvolume\rangle=0.5625$)
and
all other simulation parameters were the same as before.
The median and semiquartile range of the
effective-background area $(\mybeta)$
of these observations was, respectively,
28.04 and 4.77 $\mypx^2\,$.

\begin{center}
\begin{figure}
\includegraphics[scale=0.70,trim= 10 175 360 110]{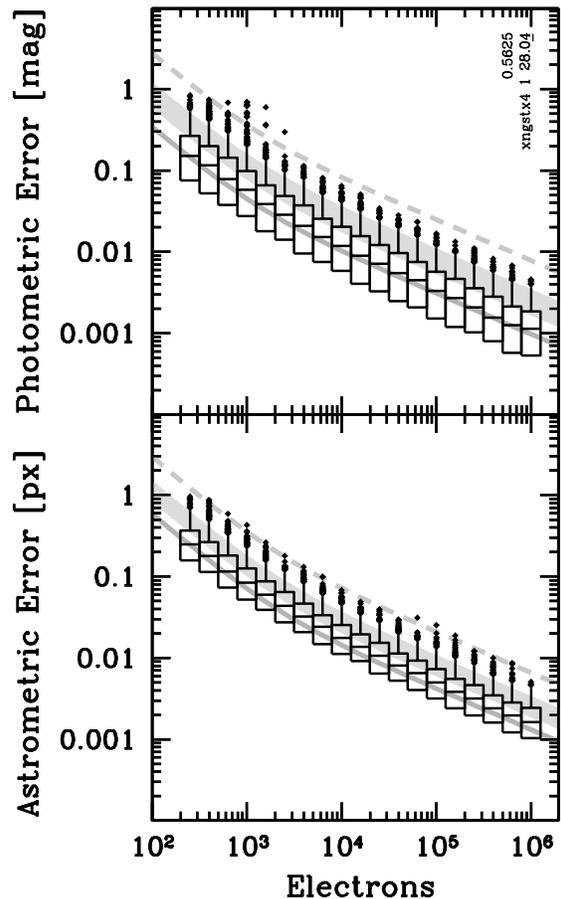}
\caption{The absolute photometric errors ({\em{top}})
and total astrometric errors ({\em{bottom}})
of $20,000$ simulated CCD stellar observations analyzed with $\mympdx$
using a 4$\times$4 supersampled version of the {\em{NGST}} PSF
$(\mybeta\approx 28.04$ $\mypx^2; \myvolume\!=\!0.5625)$.
\label{fig21}
}
\end{figure}
\end{center}

Figure~\ref{fig21} shows the
absolute photometric errors
and total astrometric errors of this numerical experiment.
Comparing the simulation results with the
grey theoretical limits, one sees that the
photometric and astrometric performance of the $\mympdx$ code is
well predicted by the theoretical performance models given in \S 2.

\begin{itemize}
\item
Excellent stellar photometry and astrometry is possible with
ugly PSFs imaged onto ugly detectors as long as the image formation process
{\em{within the detector}} is accurately modeled by the photometric
reduction code.
\end{itemize}

\section{Discussion}

After developing theoretical photometric and astrometric performance model for
Point Spread Function (PSF)-fitting stellar photometry, I
described the unique features of the MATPHOT algorithm for
accurate and precise stellar photometry and astrometry using
discrete Point Spread Functions.
I conducted numerical experiments with the $\mympd$ implementation of the
MATPHOT algorithm and demonstrated that
the computational noise due to the chosen analysis method
(numerical versus analytical)
is insignificant when compared to the unavoidable photon noise
due to the random arrival photons in any astronomical
CCD observation.
The MATPHOT algorithm was specifically designed for use with
space-based stellar observations where
PSFs of space-based cameras frequently have significant amounts of power at
higher spatial frequencies.
Using simulated {\sl{NGST}} CCD observations,
I demonstrated millipixel relative astrometry and
millimag photometry is
possible with very complicated space-based discrete PSFs.

The careful reader will observe that I have not discussed {\em{how}} a
discrete PSF is derived.  The MATPHOT algorithm will optimally determine
the brightness and position of a star in a CCD observation when provided
with the correct PSF and DRF --- functions which need to be determined
{\em{beforehand}} through calibration procedures.  Photometric and
astrometric accuracy and precision degrades
if either the PSF or DRF is poorly known.
PSF reconstruction (calibration) is a complicated topic in its own
right and has been the subject of many articles, instrumentation reports,
and entire workshops.
The challenges of PSF reconstruction are many.  An astronomer may
be faced with trying to derive a PSF from an observation
\begin{itemize}
\item
with a variable PSF within the field of view,
\item
that has too few bright stars,
\item
that might be undersampled,
\item
that might be poorly dithered,
\item
that might be poorly flat-fielded,
\item
that exhibits significant charge transfer efficiency variations,
\item
that has variable charge diffusion within the CCD substrate,
\item
with significant photon loss due to charge leakage,
\item
that might not actually be linear below the 1\% level.
\end{itemize}
While many of these problems can be overcome by the proper design of
instruments or experiments, their solution is beyond the scope of this
article which has sought to determine the practical
limits of PSF-fitting stellar photometry.

The analysis presented in this article has assumed that PSFs are perfectly
known -- a situation that is rarely, if ever, physically possible.
The {\em{cores}} of observationally based PSFs are
generally much better determined than the broad {\em{wings}} due to
simple photon statistics.
The effect of large instrumental calibration errors can also be significant.
For example,
flat-field limitations can dramatically impact the achievable
levels of photometric and astrometric precision.
An investigation based on theory of the effect of PSF errors and flat-field
calibration error on the limits of PSF-fitting stellar photometry would be
very difficult.
An investigation based on numerical experiments, however, might be a
much more tractable proposition.  In any case, a through investigation of the
effects of calibration errors on the limits of PSF-fitting stellar photometry
is best left to another article.

\vspace*{2em}
I would like to thank the anonymous referee whose careful reading and
insightful comments have improved this article.
Special thanks are due to
Ian Roederer who checked all of the
mathematical proofs in an early draft of the manuscript.
Jessica Moy has my heartfelt thanks for cheerfully helping me
acquire copies of many articles in technical journals
that are not available in the NOAO library collection.
I would also like to thank the following people
for providing stimulating conversations, encouragement, and support
during the extensive research and development effort behind the creation
of the MATPHOT algorithm and
and its various software implementations:
Joseph Bredekamp,
Susan Hoban,
Dot Appleman,
Taft Armandroff,
Nick Buchholz,
Marc Buie,
Harvey Butcher,
Julian Christou,
Chuck Claver,
Lindsey Davis,
Michele De Le Pe\~{n}a,
Mike Fitzpatrick,
Ken Freeman,
Dan Golombek,
Richard Green,
Steve Howell,
George Jacoby,
Buell Jannuzi,
Stuart Jefferies,
Ivan King,
John Krist,
Todd Lauer,
John MacKenty,
John Mather,
Mike Merrill,
Dave Monet,
Mark Morris,
Jeremy Mould,
Jan Noordam,
John Norris,
Sudhakar Prasad,
the late Alex Rogers,
Steve Ridgway,
Pat Seitzer,
James Schombert,
Donald West,
and
Sidney Wolff.
This research has been supported in part by the following
organizations,
institutions,
and
research grants
(in reverse chronological order):
National Aeronautics and Space Administration (NASA),
Interagency Order No.\ S-13811-G, which was
  awarded by the Applied Information Systems Research Program
  (NRA 01-OSS-01) of NASA's Science Mission Directorate,
Interagency Order No. S-67046-F which was awarded the
  Long-Term Space Astrophysics Program
  (NRA 95-OSS-16) of NASA's Office of Space Science,
Center for Adaptive Optics (CfAO),
Institute for Pure and Applied Mathematics (IPAM),
Kitt Peak National Observatory,
National Optical Astronomy Observatory,
Association of Universities for Research in Astronomy Inc.\ (AURA),
National Science Foundation,
Mount Stromlo and Siding Spring Observatories,
Australian National University,
the Netherlands Foundation for Astronomical Research (ASTRON),
and
the Netherlands Organization for the Advancement of Pure Research (ZWO),
and the
Kapteyn Astronomical Institute of Groningen.

\appendix
\section[]{Box-and-Whisker Plots}
\label{appendix:a}

\begin{center}
\begin{figure}
\includegraphics[trim=0 150 0 200,scale=0.45]{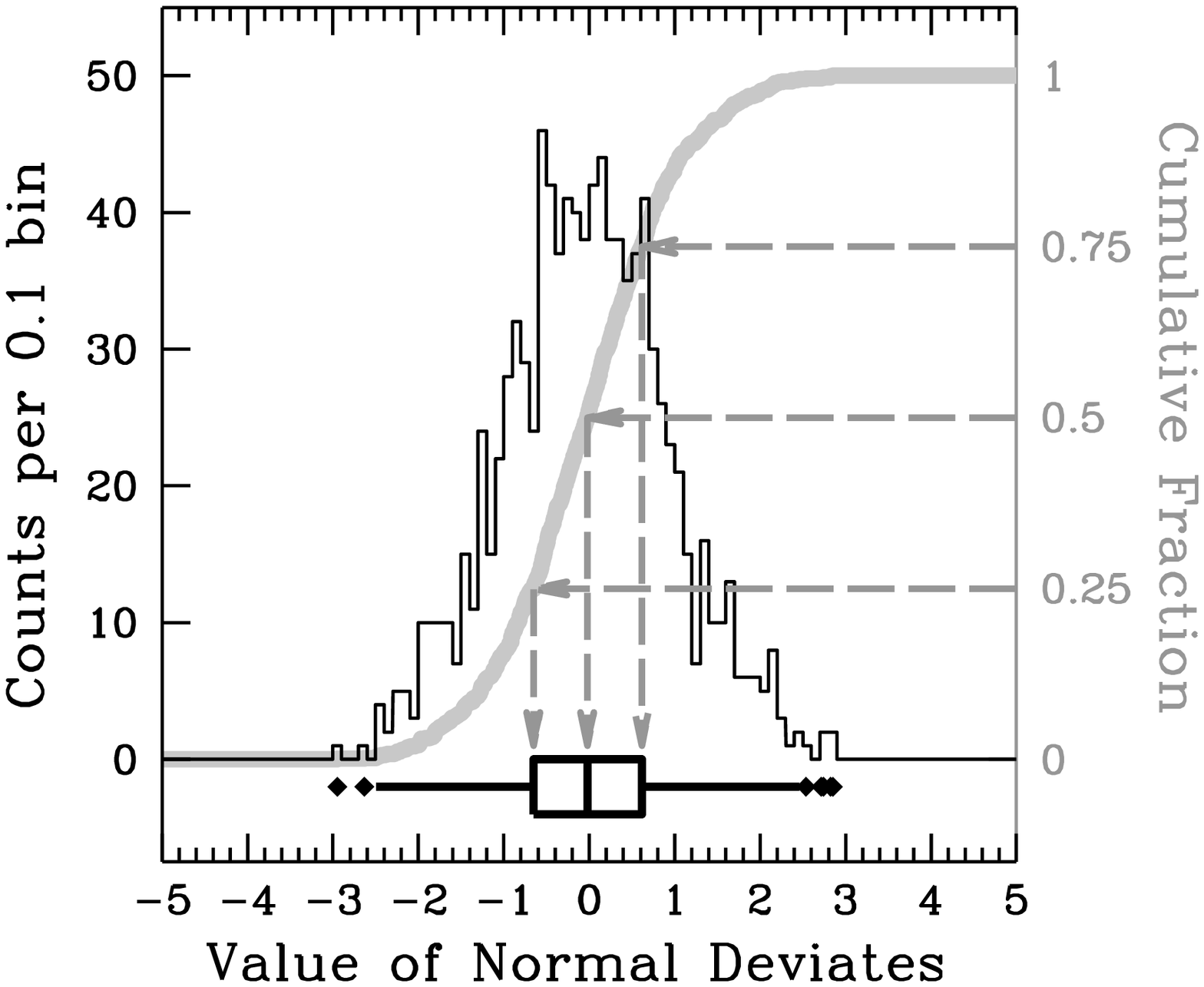}
\caption{A box-and-whiskers plot of a data set of 1000 normal deviates.
See the text for details.}
\label{figA1}
\end{figure}
\end{center}

A box-and-whisker plot (a.k.a. {\em{box plot}})
is a graphical method of showing a data distribution.
A box is drawn showing the inner quartile range of the data
which, by definition,  includes half of all the data values
(see Fig.~\ref{figA1}).
The {\em{median}} of the data is shown with a bar inside the box.
The bottom end of the box is the lower quartile (25\%) of the data;
\cite{tukey:1977}, the creator of the box-and-whiskers plot,
calls this value the {\em{lower hinge}}, $\mylh$, value.
The top end of the box is the upper quartile (75\%) of the data
which is called the {\em{upper hinge}}, $\myuh$, value.
The {\em{step}} value is 1.5 times the inner quartile range:
$\mystep \equiv 1.5*(\myuh - \mylh)$.
The {\em{top fence}} value is the sum of the upper hinge and step values:
$\mytf \equiv \myuh + \mystep$.
The {\em{bottom fence}} value is the difference between the lower hinge
and step values:
$\mybf \equiv \mylh - \mystep$.
The {\em{top whisker}} is drawn from the upper hinge value to the largest
data value that is less than or equal to the top fence value:
$\mytw \leq \mytf$.
Similarly,
the {\em{bottom whisker}} is drawn from the lower hinge value to the smallest
data value that is greater than or equal to the bottom fence value:
$\mybw \geq \mybf$.
Data values that are greater than the top fence value
or less than the bottom fence value are called {\em{outliers}}
and are plotted at their appropriate value
beyond the whiskers.
For a normal distribution,
which is a Gaussian distribution with a mean of zero
and a standard deviation of one,
the bottom fence, bottom hinge, median,
top hinge and top fence values are, respectively,
$-$2.6980 (0.35\% cumulative fraction),
$-$0.6745 (25\%),
zero (50\%),
0.6745 (75\%),
2.6980 (99.65\%).

Figure~\protect\ref{figA1} shows a data set of 1000 normal deviates.
The histogram of the data with 0.1-wide bins is shown
with thin black lines.  The cumulative fraction distribution of the data
is shown as a thick gray curve.  The box-and-whisker plot of the data
is shown with thick black lines below the histogram;
arrows show the relationship between various
box values and the cumulative fraction distribution.
The mean and standard deviation of this data set are,
$-$0.0341 and 0.9739, respectively.
The bottom fence, bottom whisker, bottom hinge, median,
top hinge, top whisker, and top fence values of this data set are,
respectively,
$-$2.5511 (0.25\% cumulative fraction),
$-$2.4940 (0.30\%),
$-$0.6522 (25.10\%),
$-$0.0231 (50.00\%),
0.6137 (75.10\%),
2.4580 (99.50\%),
2.5126 (99.57\%).
The seven outlier values of this data set,
$-$2.9500,
$-$2.6320,
2.5390,
2.7150,
2.7430,
2.8270,
2.8530,
are plotted in Fig.~\protect\ref{figA1} as diamonds beyond the whiskers.

\label{lastpage}

\clearpage

\end{document}